\documentclass[aps,twocolumn,showpacs,amsfonts,eqsecnum]{revtex4}
\usepackage{epsfig,amsmath}
\usepackage{bm}
\newcommand{\Ref}[1]{(\ref{eq:#1})}%%  requires \Ref{label}
\newcommand{\REF}[1]{Eq.~(\ref{eq:#1})}%%  requires \Ref{label}
\newcommand{\BE}[1]{\begin{equation}\label{eq:#1}}    %% Equation
\newcommand{\EE}{\end{equation}}%%   environment with label eq:#1
\newcommand{\BEA}[1]{\begin{eqnarray} \label{eq:#1}} %%  The same
\newcommand{\EEA}{\end{eqnarray}} %%           for Equation Array
\newcommand{\BEa}{\begin{eqnarray*}} %%              Not numbered
\newcommand{\EEa}{\end{eqnarray*}}   %%            Equation Array
\newcommand{\be}{\begin{equation}}
\newcommand{\bea}{\begin{eqnarray}}
 \newcommand{\BSE}[1]{\begin{subequations}\label{eq:#1}}
  \newcommand{\bse}{\begin{subequations}}
\newcommand{\ESE}{\end{subequations}}
\def\nn{\nonumber}
\def\br{\\ \nonumber}
\newcommand{\Onecol} {\begin{widetext} \onecolumngrid} %% 2 -> 1
\newcommand{\Twocol} {\end{widetext} \twocolumngrid} %% 1 -> 2
\newcommand{\B}[1]{{\bm{#1}}}%% Bold Roman & Greek Lower & Upper Case
\newcommand{\C}[1]{{\mathcal{#1}}} %% Calligrapfic Upper case
\newcommand{\BC}[1]{\bm{\mathcal{#1}}}%% Bold Calligrapfic Upper case
\newcommand{\F}[1]{{\mathfrak{#1}}}%% Fractur (Gothic) Lower & Uppers
\def\k{\B {k}} \def\r{\B {r}} %% Bold k & r
 \newcommand{\ve}{\varepsilon} %% \ell & \varepsilon
\def\<{\left \langle} \def\>{\right\rangle}%% Adjustable < & >
\def\Fbox#1{\vskip1ex\hbox to 8.5cm{\hfil\fboxsep0.3cm\fbox{%
      \parbox{8.0cm}{#1}}\hfil}\vskip1ex\noindent} %% {TEXT} in BOX
\def\Re{${\C R}\mkern-3.1mu e$} %% Reynolds number Re text mode
\def\RE{{\C R}\mkern-3.1mu e} %% Reynolds number Re, math mode
\renewcommand{\sb}[1]{_{\text {#1}}} %% sub- for lower case
\renewcommand{\sp}[1]{^{\text {#1}}} %% super- for lower case
\newcommand{\Sb}[1]{_{_{\text {#1}}}} %% Sub- for Upper case
\newcommand{\Sp}[1]{^{^{\text {#1}}}} %% Super- for Upper case
\renewcommand{\~}[1] {\tilde{#1}}
\newcommand{\PM}{$P\!M$}
\newcommand{\SPM}{_{_{P\!M}}}

\newcommand{\x}{\hat{\bf x}}

\begin{document}%%
\title{
Multi-Zone Shell Model for Turbulent Wall Bounded   Flows}

\author{Victor S. L'vov, Anna Pomyalov and Vasil Tiberkevich}

\affiliation{~Department of Chemical Physics, The Weizmann
Institute of Science,  Rehovot 76100, Israel}

\begin{abstract}
We suggested  a \emph{Multi-Zone Shell} (MZS)  model  for
wall-bounded flows  accounting for the space inhomogeneity  in a
``piecewise approximation", in which  cross-section area of the
flow, $S$, is subdivided into ``$j$-zones". The area of the first
zone, responsible for the core of the flow, $S_1\simeq S/2$, and
areas of the next $j$-zones, $S_j$,  decrease towards the wall
like $S_j\propto 2^{-j}$. In each $j$-zone the statistics of
turbulence is assumed to be space homogeneous and is described by
the set of ``shell velocities" $u_{nj}(t)$ for turbulent
fluctuations of the scale $\propto 2^{-n}$. The  MZS-model
includes a new set of complex variables, $V_j(t)$, $j=1,2,\dots
\infty$, describing the amplitudes of the near wall coherent
structures  of the scale $s_j\sim 2^{-j}$ and responsible for the
mean velocity profile. Suggested MZS-equations of motion for
$u_{nj}(t)$ and $V_j(t)$ preserve the actual conservations laws
(energy, mechanical and angular momenta), respect the existing
symmetries (including Galilean and scale invariance) and account
for the type of the non-linearity in the Navier-Stokes equation,
dimensional reasoning, etc. The  MZS-model   qualitatively
describes important characteristics of the  wall bounded
turbulence, e.g., evolution of the mean velocity profile with
increasing Reynolds number, $\RE$, from the laminar profile
towards the universal logarithmic profile near the flat-plane
boundary layer as $\RE\to \infty$.
\end{abstract}%%
\maketitle%%

\section*{\label{s:intro}INTRODUCTION}%%

\subsection{\label{ss:background}Background}%%

Three simple turbulent flows -- in a channel, in a pipe and near a
flat plane -- play a prominent role in our understanding of
spatially inhomogeneous wall bounded flows, similar to the
celebrative role of the developed homogeneous turbulence in the
understanding the universal statistical behavior of fine-scale
turbulence. On the long way of understanding homogeneous
turbulence there appeared various phenomenological cascade models
of turbulence (Richardson-Kolmogorov-41 concept of turbulence,
Kolmogorov-62 log-normal and ``multi-fractal" models of
intermittency), many closure procedures (like Kraichnan Direct
Interaction Approximation), various field theoretical approaches.
Last but not least we have to mention so-called ``shell models" of
turbulence (like GOY shell model~\cite{Gledzer,GOY} together with
its "Sabra" improvement~\cite{98LPPPV}, and many
others~\cite{shell-models}). Separately we want to mention Zimin's
shell model (V. Zimin, private communication, see
also~\cite{95ZF,93FZ}) which has been derived from the
Navier-Stokes equation (NSE) using a vector wavelet
decomposition~\cite{91Men} of the velocity field, and which
involves no empirical or ad hoc parameters.

The shell models are systems of ordinary differential equations
which mimic the statistically homogeneous isotropic turbulent
velocity field in some interval of scales (say, within some
``shell" in the Fourier space) by one or few ``shell velocities"
$u_n(t)$~\cite{shell-models}. The shell models have the same
(quadratic) type of nonlinearity, as the NSE, respect the
conservation of energy (in unforced, inviscid limit) and have the
build-in ``locality" of interaction of neighboring scales,
reflecting scale-by-scale energy transfer toward dissipative
scales. Surprisingly, the shell models allows to mimic almost
everything we know (experimentally, theoretically or by direct
numerical simulation) about highly nontrivial statistics of
fine-scale turbulence. This include, for instance, the
``intermittent" behavior of the velocity structure functions
(which are simultaneous, two-point $n$th-order correlation
function of velocity differences), the fusion rules (which govern
asymptotic behavior of many-point velocity correlation functions),
and so on. A possible reason for such a lucky success is that the
above mentioned (and some other) characteristics of the turbulent
statistics are robust and depend only on some very general
physical requirements, such as respect of the actual conservation
laws , the scale invariance, the type of nonlinearity in the NSE,
and the locality of interaction. All these features are accounted
for in the shell models.

Unfortunately, the shell models in their traditional formulations
describe only the space homogeneous turbulence, leaving aside the
wall bounded turbulence, which play a much more important role in
practical applications.

Turbulent flows at high Reynolds numbers, $\RE\gg 1$, contain such
a wide range of excited lengths and time scales that the direct
numerical simulations (DNS) of the NSE are impossible for the
foreseeable future. Consequently, practical engineering
calculations are based on some model simplifications of NSE,  with
Reynolds stress models being the most popular approach, see e.g.
the book \cite{Pop}, review~\cite{78Lum} and references therein.
The idea of Osborne Reynolds (see, e.g. Ref.~\cite{MY}) was to
divide the velocity field into the mean flow part $\B V(\r)$ and
the turbulent fluctuating part $\B u(\B r,t)$ with zero mean and
to approximate in some way the hierarchy of equations for various
correlation functions (correlators). Equation for $\B V(\r)$
contains so-called Reynolds stress term, a second order correlator
of $\B u(\B r,t)$. The right hand side (RHS) of equation for the
Reynolds stress contains five different terms: the rates of
production, dissipation, turbulent transport and viscous
diffusion, and the velocity pressure gradient term. These are
one-point, second and third order correlators of velocity and
velocity gradients and the pressure-gradient correlator. The
equation for only one such object, the dissipation rate, already
contains eight correlators up to the 4th order in velocity, which
are usually modeled by various closure procedures in terms of the
lower order objects. The simplest, old-fashion Millionshchikov's
closure (Monin and Yaglom,~\cite{MY}, p. 241) is often invoked. It
uses a non-realistic assumption of Gaussian statistics of
turbulence. To improve the situation one can use a set of
phenomenological constants, which can be found by comparison of
the results of model calculations with the results of DNS or
experiments in the benchmark flows.

There where  attempts  to use more advanced field theoretical
approaches developed in the theory of homogeneous turbulence, e.g.
the Yakhot-Orszag  version of the Renormalization Group (RNG)
approach~\cite{86YO,92YOTGS}. Instead of going into detailed
criticism of this approach, already done in Ref.~\cite{92SR}, we
just make two general remarks to the RNG approach, that are also
relevant to the most attempts of a straightforward transfer of the
field theoretical methods of fully developed, fine scale,
homogeneous turbulence to  the case of wall bounded flows. First,
in  most cases, (RNG, diagrammatic perturbation approach, etc.)
the turbulence is assumed to be excited by some artificial
external force with Gaussian statistics. This is reasonable
simplification of the real picture, if one deals with turbulent
scales that are deep enough in the inertial interval. However,
this is definitely not a realistic assumption for large scales,
that are important in the energy and mechanical momentum balance
in wall bounded turbulence. Second, in  most cases, the
field-theoretical approaches to homogeneous turbulence are usually
formulated in the $\k$-representation or explicitly assumed the
space homogeneity. In this way one gets required closure
relationships, say between the effective turbulent viscosity
$\nu\Sb T$, the density of the kinetic energy $\C E$ and the rate
of energy dissipation (used in the popular
$\overline{K}$-$\overline{\epsilon}$ version of the Reynolds
stress model). However, at least two of these objects ($\nu\Sb T$
and $\C E$) are not locally defined, they are dominated by the
largest eddies in the system, usually of scales close to the
distance to the wall. Therefore one has to be extremely careful
applying the resulting relations to the wall bounded flows in
which the characteristic length of inhomogeneity is exactly the
distance to the wall. A price to pay for such a simplification is
that the phenomenological constants may depend on the flow
geometry or even on the position in the flow.

Introducing enough adjustable parameters (sometimes geometry
dependent), one can reach an engineering goal to model in computer
some mean and turbulent characteristics of particular flows of
practical importance. However, an important aspects of the basic
physics of wall bounded turbulence remain unclear,  being masked
by numerous details, or even incorrectly reproduced.

The main goal of this paper is to suggest a physically transparent
and analytically analyzable model of wall bounded flows, oriented
mainly  on physical community. The model describes an interplay of
two main physical phenomena in the wall flows: the energy cascade
toward small scale (like in the developed homogeneous turbulence)
and the cascade of the mechanical momentum toward the wall in the
physical space. Our model is  a generalization of  the shell model
of homogeneous turbulence on the case of inhomogeneous turbulence
and accounts for a non-uniform profile of the mean velocity.
Simplifying assumptions are made from the very beginning, at the
level of basic, dynamical equations of motion. These equations
involve only two parameters, responsible for the energy and
mechanical momentum fluxes. These parameters can be evaluated from
the NSE, but currently they are chosen to reproduce the Von Karman
constant $\kappa\Sb K $ and constant $B$ in the universal
log-profile of the mean velocity near flat plane.

Our model is oriented toward the classical examples of the wall
turbulence, like the channel and the pipe flows, the planar and
the circular Couette flows. The physical description of the model
and its equations of motion are presented in the following
subsection.%%

\subsection{\label{ss:MSM-brief}%%
Multi-zone Shell Model in brief}
 For dealing with inhomogeneous wall turbulence we suggest in this
paper a \emph{piecewise homogeneity approximation}  in which the
cross-section area of the flow, $S$, is subdivided into a set of
\emph{``$j$-zones"}. The area of the first zone, responsible for
the core of the flow, $S_1\approx S/2$, and the areas of the next
$j$-zones, $S_j$, decrease towards the wall like $S_j\propto
2^{-j}$. In each $j$-zone, the statistics of turbulence is assumed
to be space homogeneous and is described by its own shell model%%
\bea \label{eq:u-nj} %%
\frac{d u_{nj}(t)}{d \, t} &\!\!\!=\!\!\!&  - \nu_n\, \kappa_{n}^2
u_{nj} + \C N_{nj}+\Delta_{nj}\,  w_{n}\,,%%
\EEA%%
for  the ``shell velocities" $u_{nj}(t)$, which are responsible
for turbulent fluctuations of the (dimensionless) scale $s_n \sim
2^{-n}$, referred for  below  as  $(nj)$-\emph{eddies}. Hereafter
$\Delta_{nj}$ is the Kronicker symbol ($=1$ for $n=j$ and $=0$
otherwise). Equation~\Ref{u-nj} accounts for the viscous damping
term with some effective viscosity $ \nu_n \sim \nu_0$, where
$\nu_0$ is  the kinematic viscosity of the fluid. The effective
``shell" wave vector $\kappa_n\propto 1/s_n$. The nonlinear term
in~\REF{u-nj}, $\C N_{nj}$, is given by \REF{N-gen1} and describes
the usual triad interaction of nearest shells, inside of a given
$j$-zone, \REF{Nnj-Sabra},  and some inter-zone interaction term
of similar type. The production term $w_n$ is responsible for the
energy flux from the mean flow to the turbulent subsystem and is
given below by \REF{w-n}.

Our goal is to describe the mean velocity profile $\< \B V(\B
\rho,t)\>$, in which $\B \rho$ is two dimensional radius-vector in
the cross-section of the flow, $\B \rho\bot \hat{\bf x}$,
$\hat{\bf x}$ is  the streamwise direction. To this end we
introduce additional variables $V_j(t)$ with a prescribed space
dependence $\B \Phi_j(\B \rho)$, uniquely determined by the flow
geometry. The variables $V_j(t)$ can be understood as  complex
amplitudes of the near-wall coherent structures  of the
dimensionless scale $s_j$, which is the same, as the scale of
$(nj)$-eddies: $s_j=h_n$ for $n=j$. The functions $\B \Phi_j(\B
\rho)$ are chosen such that $\B \Phi_{j+1} (\B \rho)\simeq \B
\Phi_j(2 \B \rho )$ and form an orthonormal, (but incomplete)
basis. We call it ``\emph{PM-basis}", because (in spite of its
incompleteness) it is chosen such as to represent \emph{exactly}
the densities of the mechanical linear and angular momenta, $\C P$
and $\C M$, in terms of $V_j(t)$ only:%%
\bea\label{eq:lin-mom} %%
\C P = \sum_j s_j \, \text{Re}\big[V_j(t)\big]\,,
 \  \C M = \sum_j s_j R_j \,\text{Im} \big[V_j(t)\big]\ .%%
\EEA %%
Here $R_j$ is the characteristic distance of the $j$-zone from the
centerline of the flow. The $P\!M$-basis also allows one to
reconstruct the spatial dependence of the mean flow (and its time
depending fluctuations) with a finite (but very good) accuracy%%
\bea\label{eq:rec}%%
\B V (\B \rho,t)&\approx& \B V\Sb{PM} (\B \rho,t)\equiv
\text{Re}\Big[\sum_j V_j(t)\,\B \Phi_j(\B \rho)\Big]\ .%%
\end{eqnarray}%%

For the \emph{$P\!M$-velocities} $V_j(t)$ we suggest a simple
\emph{momentum equation}%%
\bea\label{eq:V-n}%%
\frac{d V_j(t)}{d t}=-\nu_j \kappa_j^2 V_j + \nabla p +
  W_j\,,
\EEA%%%
which includes  the viscous term,   $\nu_j \kappa_j^2 V_j$,
pressure gradient, $\nabla p >0$,  and the Reynolds stress term,
$W_j$, that   accounts for the exchange of the mechanical momentum
between nearest zones in the flow and is given by \REF{W-n}.

It is crucially important that suggested  \emph{multi-zone shell
model}, Eqs. \Ref{u-nj} and \Ref{V-n},  preserves (in the
unforced, inviscid limit)  all the relevant in the problem
integrals of motion: energy, $\C E$, mechanical  momenta, $\C P$
and $\C M$.

\subsection{\label{ss:plan}The plan of the paper}
Section~\ref{s:multizone} is devoted to the  statistical
description of the ``turbulent part" of the Multi-Zone Shell model
(MZS-model), \REF{u-nj}. First, in Sec.~\ref{ss:NSE2shell} we
describe a way from the NSE to a standard shell model of
homogeneous turbulence, that allows generalization for space
inhomogeneous turbulence. Next, in Sec.~\ref{ss:PHA} we formulate
a piece-wise homogeneity approximation: the cross-section area of
the flow, $S$, is subdivided into set of ``$j$-zones", in each of
them  the statistics of turbulence is assumed to be space
homogeneous. This allows us to use the standard shell model for
every $j$-zone and to describe the turbulence in the whole flow by
the set of ``shell velocities" $u_{nj}(t)$ for turbulent
fluctuations of the scale $\propto 1/2^n$.

In Sec.~\ref{s:moment-eq} we derive the dynamical equation of
motion for the so-called \PM-velocities, $V_j$,  that allows one
to reconstruct with good accuracy the mean velocity profile $\B
V(\B \rho)$ in the cross-section of the flow. In particular, in
Sec.~\ref{ss:PM-basis} we introduce the \PM-basis for a wide class
of wall bounded flows that connects $\B V(\B \rho)$ and $V_j$. In
Sec.~\ref{ss:simple-Vj-eq} we derive \REF{V-n} for $V_j$ and for
all involved in \REF{V-n} terms.

In Section~\ref{s:analysis} we discuss the conservation laws
(Sec.~\ref{s:fluxes}) and the symmetries (Sects.~\ref{ss:Galilei}
and \ref{ss:as-univ}) of the MZS model. For the convenience of the
reader it begins with the overview of the resulting MZS equations,
presenting them, in Sec.~\ref{ss:res}, in the dimensionless form
\Ref{MSM}, convenient for the further analysis. In particular, we
derive in Sec.~\ref{ss:as-univ} the version of the MZS model
 for the turbulent boundary layer near flat plane, Eqs.~\Ref{rMSM}.

Section~\ref{s:sol} presents a detailed analytical study of the
MZS equations in a set of approximations, realistic at various
values of $\RE$. In Sec.~\ref{ss:laminar} we show that the
MZS-model describes the stable laminar velocity profile for small
$\RE$ and  its instability at some threshold value of
$\RE=\RE\sb{cr}$. Next, in Sec.~\ref{ss:abc=0} we study the
MZS-model in the approximation of the near-wall eddies, neglecting
the turbulent energy cascade. This effect is accounted  for in
Sec.~\ref{ss:log} in the approximation of turbulent viscosity.
Section~\ref{ss:numerics} is devoted to numerical analysis of the
MZS-model.

In Section~\ref{s:sum}, we summarize our findings and suggest  a
possible  generalization of the  model for the description of
turbulent flows laden with the long-chain polymeric additives (in
connection with the problem of drag reduction) or with heavy
micro-particles, etc.

\subsubsection{\label{ss:nom}Some notations and definitions}

\textbullet~~$\x$ \& $\B \rho$ -- the streamwise direction \& the
two-dimensional radius-vector in the cross-section of a flow, $\B
\rho\bot \x $. In a channel $\B \rho= (y, \, z)$, with $y$ as the
wall-normal and $z$ as the spanwise directions. %%%%

\textbullet~~$S_\bot$, $P_\bot$ \&  $L$ --  the cross-section
area,  the perimeter and  the characteristic length of a
cross-section:%%
\bea\label{eq:def-L}%%
S_\bot=\int d\B \rho\,, \quad    L\equiv S_\bot/P_\bot\ .%%
\EEA%%
In a channel of width $2H$, $L=H$; in a pipe of radius $R$,
$L=R/2$.

\textbullet~~$\nabla p$ --  the external pressure gradient, that
is a positive constant %%
\bea\label{eq:def-np}%%
\nabla p\equiv - \frac{d p(x)}{dx}> 0 \ .\EEA %%

\textbullet~~$\tau$ \& $U_\tau$ -- the characteristic time \& the
velocity in the flow:%%
\bea\label{eq:def-TV}%%
\tau\equiv \sqrt{L/\nabla p}\,, \qquad U_\tau\equiv \sqrt {L
\nabla p}\ .\EEA %%
The wall shear stress is $U_\tau^2$.

\textbullet~~$\nu_0$ \& $\RE$ -- the kinematic viscosity \&
 the friction Reynolds number%%
\bea\label{eq:def-RE} %%
\RE\equiv U_\tau L/\nu_0 \ .%%
\EEA %%

\textbullet~~$f'$ \& $f''$ --  the real and imaginary parts of
some complex object $f$ (constant, variable, function, etc.):%%
\bea\label{eq:IR}%%
 f' \equiv \text{Re}\{f\}\,, \qquad f''\equiv
\text{Im}\{f\} \ .%%
\EEA%%

\textbullet~~$n$, $j$ \& $p$ are dummy indices (natural numbers),
reserved for the scale (or shell), zone \& position indices; some
objects can be related both to the shells and zones. They  are
used with both $n$ or $j$ indices.

\textbullet~~$s_j$ \& $\sigma_j$  -- the fraction of the
cross-section area, occupied by the $j$-zone, \& the fraction of
the cross-section area  occupied by all zones towards the wall,
starting  from the zone $j$:%%
\bea\label{eq:def-s}%%
s_j\equiv \frac{S_j}{S_\bot} \,, \quad    \sigma _j\equiv
\sum_{i=j}^\infty s_i\,, \quad \sigma_1=
\sum_{j=1}^\infty s_j=1\ . %%%%
\EEA %%

\textbullet~~$L_j\equiv s_j \, L$ -- the  width of the $j$-zone,
$L_n$ -- the characteristic scale in the $n$-shell%%

\noindent\textbullet~~$\kappa_n \equiv  1/(2\,L_n)$ -- the wave
number in the $n$-shell; $\kappa_j$ -- the wave number of the
$j$-zone

\textbullet~~$\C E$, $\C P$,  \& $\C M$ -- the densities of the
energy, of the mechanical linear \& angular momenta ($\x$
projections)

\textbullet~~$\ve ^\pm\ $,  $\F p^\pm,\ $ \& $~\F m^\pm$ -- the
total rates of pumping (with $^+$) and dissipation  (with $^-$) of
the conserved quantities  $\C E$, $\C P$,  \& $\C M$

\textbullet~~$\ve ^\pm_j\ $,  $\F p^\pm_j,\ $ \& $~\F m^\pm_j $ --
the rates of pumping (with $^+$) and dissipation (with $^-$) in
the $j$-zone (of $\C E$, $\C P$, \& $\C M$)

\textbullet~~$\ve_j\ $,  $\F p_j,\ $ \& $~\F m _j $ -- the fluxes
of $\C E$, $\C P$, \& $\C M$  from the $j$- to the $(j+1)$-zone

\textbullet~~The scalar product of complex vector functions, $\B
A(\B \rho)$ \& $\B B(\B \rho)$: %%
\bea \label{eq:def-AB}%%
 (\B A,\B B)&\equiv&
\int \B A^*(\B \rho)\cdot \B B(\B \rho)
  \frac {d\B \rho}{S_\perp}\ .%%%%
\EEA%%

\textbullet~~$\B \phi^+_m(\B \rho) $ \& $\B \phi^-_m(\B \rho) $
the even and odd eigenfunctions of the two-dimensional Laplace
operator in the cross-section of the flow with no-slip boundary
conditions, $\B \phi^\pm _m(-\B \rho)=\pm \B \phi^\pm(\B \rho)$%%

 \textbullet~~$\B \Phi_j(\B \rho )= \B \Phi_j'(\B \rho )+
 i \B \Phi_j''(\B \rho )$ -- \PM-basis functions, \REF{Phi-n}

\textbullet~~$\B V\SPM (\B \rho)$ \& $V_j$ -- \PM-velocity in the
coordinate and $j$-representations, related by \REF{expV-PM} %%%%

\textbullet~~$u_{nj}$ -- the velocity of turbulent fluctuation of
the scale $\propto 2^{-n}$ in the $j$-zone [$(nj)$-eddies];
$u_j\equiv u_{jj}$%%

\section{\label{s:multizone}%%
Statistical Multi-zone shell model  for turbulent fluctuations}%%
\subsection{\label{ss:NSE2shell}%%
From  NSE to shell models of homogeneous turbulence}%%
In this Subsection we present a re-derivation  of a standard shell
model of space homogeneous turbulence in a way that allows us to
generalize it in Sec.~\ref{ss:PHA} for the case of space
inhomogeneity.%%

\subsubsection{``Cell basis", wavelets and ``\,$(n\B p)$-eddies"} %%
Consider for simplicity an incompressible turbulent velocity $\B
u(\r,t)$ in a periodic box of size $L\times L\times L$. Instead of
the $\r$- or $\k$-representation we introduce here a ``\emph{cell
basis}" $\B \Psi_{n\B p}(\r)$,  which is  quite similar to wavelet
bases (for an easy-to-read, introductory text about the theory of
wavelets see, e.g.~\cite{95Hol}). Similar to the wavelet bases,
the cell bases reflect both spatial scales of turbulent structures
(like in the $\k$-representation) and their position in the
physical space (like $\B r$-representation) but accounts for the
actual boundary conditions of the flow. \emph{The cell index},
$({n,\B p})$, consists of \emph{the scale  index}
$n=1,2,\dots\infty$ and \emph{the position index }$\B
p=(p_x,p_y,p_z)$:

\emph{The scale index } $n=1,2,\dots\infty$ defines the
characteristic width $L_n$ (in all directions)  of the function
$\B \Psi_{n\B p}(\r)$, which  some ``$(n\B p)$-cell" of the size
$L_n = L/2^n$ ``occupies" in space.

\emph{The position index }$\B p=(p_x,p_y,p_z)$ defines the
position $\B R_{n\B p}$ of the $(n\B p)$-cell:%%
\bea\label{eq:Rns}%%
\B R_{n\B p}&\!\!\!\equiv\!\!\!&\B p\,L_n\,,\quad \B p=
(\,p_x,\,p_y,\,p_z\,)\,,\br %%
\B p\in \C C_n  &\Leftrightarrow& \quad p_\alpha=1,2,\dots 2^n\,,
\qquad \alpha = x,y,z\ .%%
\EEA %%
One can imagine that, for any given  $n$, the set of  $(n\B
p)$-cells with $\B p\in \C C_n$ fills the periodical $L^3$-box.
The turbulent velocity field $\B u(\r,t)$ is  given by: %%
\bea\label{eq:shell-exp}
 \B u(\B r,t) &=& \sum_{n=1}^\infty
\sum_{\B p\in \C C_n}\text{Re}\big\{ U_{ n \B p}(t)\B \Psi_{n \B
p}(\B r)\big\}\ ,%%
\EEA %%%
where the amplitude of the ``cell expansion", $U_{n\B p}(t)$ is
the velocity difference across the separation  $L_n$ in the  $(n\B
p)$-cell.

It is convenient to normalize the cell-basis  as follows:%%
\bse\label{eq:orth-s3}\bea%%
\int \limits_0^{L}\!\!\!\!\int \limits_0^{L}
\!\!\!\!\int \limits_0^{L}
\bm\Psi_{n\bm p}^*(\r)\cdot\bm\Psi_{n'\bm p'}(\r)\frac{dxdydz}{L^3}
 &=& 2v_n\Delta_{nn'}\Delta_{\bm p\bm p'}
\,,\nonumber\\
\int \limits_0^{L}\!\!\!\!\int \limits_0^{L} \!\!\!\!\int
\limits_0^{L} \bm\Psi_{n\bm p}(\r)\cdot\bm\Psi_{n'\bm
p'}(\r)\frac{dx dy dz}{L^3}
 &=& 0
\ . \EEA\ESE%%
Here $v_n=2^{-3n}$ is the dimensionless part of the total volume
per one mode in the $n$th cell.

The equations \Ref{shell-exp} and \Ref{orth-s3} give the Parseval
identity for the density of the turbulent energy in the form: %%
\bse\label{eq:3-Pars}%%
\begin{eqnarray}\label{eq:Pars-a}
\C E&\equiv& \int \limits_0^{L}\!\!\!\!\int \limits_0^{L}
\!\!\!\!\int \limits_0^{L} \frac{|\B u(\r,t)|^2}2
\frac{dxdydz}{L^3}=
\sum_{n=1}^\infty \C E_n(t)\,,\\
\label{eq:Pars-b}%%
~\hskip -3cm
 \C E_n(t)&=& v_n\sum_{\bm p\in\C C_n}\frac{|U_{n\B p}(t)|^2}2\ .
\end{eqnarray}\ESE%%
This equation  supports our interpretation of $U_{n\B p}(t)$ as
the velocity difference across the separation $L_n$ in the $(n\B
p)$-cell.  We will refer to these fluctuations as to ``$(n\B
p)$-eddy".

A particular choice  of the cell functions $\B \Psi_{n\B p}(\r)$
is not important for us here. Notice only, that for large $n$
basic cell functions become scale invariant and may be obtained by
dilatations of one (or few) $n$-independent function
$\B\Psi_\infty(x)$. In this limit the cell basis becomes the wavelet
one with $\B\Psi_\infty(x)$ as so-called ``$\C
R$-wavelet"~\cite{95Hol}.  An explicit example of a
divergence-free 3 dimensional vector $\C R$-wavelet function given
in \cite{95ZF}.

The cell functions $\B \Psi_{n\B p}(\B r)$ form a complete
orthonormal basis, therefore one can derive the \emph{exact
equation of motion } for $U_{n\B p}(t)$ by the Galerkin projection
of NSE:
\begin{eqnarray}\label{eq:shell-eq-exact}
&& \frac{d U_{n \B p}^\sigma(t)}{d\,  t } =-\sum_{\B p'}\Gamma
_{n,\B p\B p'}U_{n \B p'}^\sigma(t)\\ \nonumber && +
\sum_{n'n^{\prime\prime}}\ \sum_{\B p'\B p^{\prime\prime}} %%
\ \sum_{\sigma' \sigma''=\pm}T_{nn'n^{\prime\prime},\B p \B p'\B
p^{\prime\prime}} ^{\sigma\sigma'\sigma''} U_{n'\B
p'}^{\sigma'} (t)U_{n^{\prime\prime}\B
p^{\prime\prime}}^{\sigma''}(t)\,,
\end{eqnarray}
where $\sigma,\sigma',\sigma''=\pm$ are sign indices and we accept
the convention $U_{n \B p}^-\equiv U_{n \B p}$ and $U_{n \B
p}^+\equiv U_{n \B p}^*$. The explicit forms of the damping
parameters $\B\Gamma$ and amplitudes $\B T$ depend on the basis,
see, e.g.~\cite{95ZF}.%%

\subsubsection{\label{ss:basic-ass}%%
Basic assumptions for the standard shell models}%%
Consider shortly the physical simplifications that allows one to
reduce the NSE to the shell model of homogeneous turbulence.
Unlike the similar discussion in~\cite{95ZF}, we made an accent on
the assumption of space homogeneity and the possibility to relax
this assumption in order to generalize shell models for the space
inhomogeneous case.

The standard shell models of homogeneous turbulence follow
 from  the exact Eq.~(\ref{eq:shell-eq-exact}) with the following
 simplifying assumption of a \emph{statistical nature}:%%
\bse\label{eq:simpl-1}\begin{eqnarray}\label{eq:simpl-1a}%%
U_{n\B p}(t)&\Rightarrow& u_{n}(t) A_{n\B p}\,, %%
\\ \label{eq:simpl-1b}%%
\overline{A_{n\B p}}&=&0\,, \qquad\overline{|A_{n\B p}|^2}=1\ .%%
\end{eqnarray}\ESE%%
Here $A_{n\B p}$ are \emph{time independent}, random amplitudes,
and \ $ ~^{\overline{~~~~ }}$~~denotes averaging over yet unknown
statistics of $A_{n\B p}$, which is generated by the NSE
\Ref{shell-eq-exact}. The physical meaning of $u_n(t)$ is the mean
square velocity \emph{fluctuations} of all $(n\B p)$-eddies :%%
\bea\label{eq:meam}%%
\<\overline{|U_{n\B p}(t)|^2}\>= \<|u_n(t)|^2\>\,, %%
\EEA%%
with the mean velocity equal to zero $\<\overline{U_{n\B
p}(t)}\>=0$, even if the time averaged shell velocity
$\<u_n(t)\>\ne 0$.

The physical arguments behind Eq.~(\ref{eq:simpl-1}) may be based
on the fact that in the homogeneous turbulence all velocities
$U_{n\B p}(t)$ with different $\B p$ have the same statistics.
Eq.~(\ref{eq:simpl-1}) therefore neglects only the difference
between the \emph{actual time realizations} of $n\B p$-velocities
$U_{n\B p}(t)$ of the same scale (the same scale index $n$), but
occupying different  cells (different position index $\B p$). The
ensemble of the time realizations is replaced by the
time-independent ensemble  of $A_{n\B p}$. With the
assumption~(\ref{eq:simpl-1}), Eq.~(\ref{eq:shell-eq-exact})
yields:%%
\bse\label{eq:shell-eq-2}%%
\begin{eqnarray}\label{eq:shell-eq-2a}
&& \hskip -2cm \frac{d u_{n}(t)}{d\,  t } = -\gamma _{n} u_{n}(t)
+ N_n\,,\\ \label{eq:shell-eq-2d}%%
&& \hskip -1.35cm \gamma_n=\sum_{\B p'}\Gamma _{n,\B p\B p'}
  \overline{A_{n\bm p}A_{n\bm p'}}
\,,\\ \label{eq:shell-eq-2b}%%
&& \hskip -1.45cm  N_n=\sum_{n' n^{\prime\prime},
\sigma'\sigma^{\prime\prime} }\! \!\! \!\!
S_{nn'n^{\prime\prime}}^{\sigma'\sigma^{\prime\prime}}
u_{n'}^{\sigma'}(t)u_{n^{\prime\prime}}^{ \sigma^{\prime\prime} }
(t)\,,
\\  \label{eq:shell-eq-2c} %%
&& \hskip -2cm
S_{nn'n^{\prime\prime}}^{\sigma'\sigma^{\prime\prime}}= \sum_{\B
p'\B p''} T_{nn'n^{\prime\prime},\B p \B p'\B
p^{\prime\prime}}^{\sigma'\sigma^{\prime\prime}}\overline{A_{n\B
p} A_{n'\B p'}A_{n''\B p''} }\  .
\end{eqnarray}\ESE%%
The correlation functions of $A_{n\B p}$ can be evaluated with
some reasonable \emph{statistical} assumptions (for instance, with
some closure procedure, like  Direct Interaction Approximation,
see e.g.~\cite{95ZF}).

Under assumption~(\ref{eq:simpl-1}), Eq.~(\ref{eq:Pars-b}) gives a
usual equation for the density of the total energy of the $n$th
scale:  $\C E_n(t)= \frac12\big|u_{n}(t)\big|^2$\ . Different
shell models correspond to various further simplifications of the
nonlinear term~(\ref{eq:shell-eq-2b}). For example, in the Sabra
shell model~\cite{98LPPPV}:%%
\bse\label{eq:Nn-Sabra}\begin{eqnarray}\nonumber
 N_n&=&i  \big[ a \kappa _{n+1}
  u_{n+1}^*u_{n+2}
  + b\,\kappa _{n} u_{n-1}^*u_{n+1} \\ \label{eq:Nn-Sabra-a}
  && - c\kappa _{n-1} u_{n-2} u_{n-1} \big]\,,\\ \label{eq:Nn-Sabra-b}
 \kappa_n&\propto & 2^n \,,\qquad a+b+c=0\ .
\end{eqnarray}\ESE%%
Notice, that the scale index $n$ in the  Eqs.~\Ref{shell-eq-2} and
\Ref{Nn-Sabra} for shell models  becomes the \emph{shell index}.

\subsection{\label{ss:PHA}%%
Piece-wise homogeneity approximation and multi-zone shell model
for turbulent fluctuations}%%
\begin{figure}
\includegraphics[width=7.5cm]{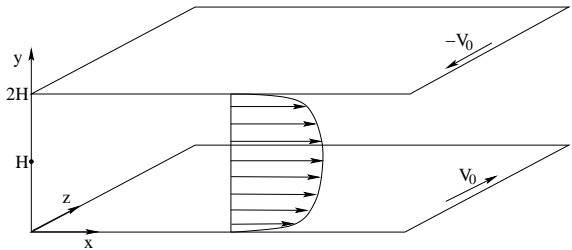}
\caption{\label{f:geometry} Geometry of the channel and the plane
Couette flow between two parallel planes separated by $2H$ in the
cross-stream direction $y$. In the simple channel flow the
pressure gradient is applied in the ``streamwise" direction $x$
and the mean velocity $\overline{\B V}(y)$ is oriented also along
$x$. In the plane Couette flow the lower wall ($y=0$) is moving in
$z$ (span-wise) direction with some velocity $V_0$, while the
upper wall ($y=2H$) is moving in the opposite direction. In this
case $\overline{\B V}(y)$ has  $z$-projection. For both flows (and
their hybridization) the three-dimensional velocity fluctuations
are space-homogeneous in the $x-z$ plane.  }
\end{figure}

The turbulent fluctuations in the wall bounded flows are not space
homogeneous due to the spatial dependence of the mean velocity
profile, which, in its turn, is also affected by the turbulent
fluctuations. Due to the inhomogeneity of turbulence, the shell
model approach discussed in Sec.~\ref{ss:NSE2shell} have to be
revised, which is the subject of this subsection. For concreteness
we discuss the planar geometry of the channel flow of width $2H$
in the cross-stream direction $y$, see Fig. \ref{f:geometry}.  In
further analysis we consider only the low half of the channel,
$0<y<H$, having in mind that the flow in the second half of the
channel, $H<y< 2\,H$, is statistically identical to the that in
the first one.
%%%%

Clearly, the core of the flow (say, for  $H/2 < y < H$) may be
approximately viewed as homogeneous. Let us call this region the
``1-zone". The next region  $H/4<y<H/2$, in which the mean
velocity profile $V(y)$ begin to decrease towards the wall, we
call the ``2-zone". Notice, that the width of the 2-zone,
$H_2=H/4$ is a half of the 1-zone width, $H_1=H/2$. Therefore,
approximately with the same accuracy we can consider the
statistics  of turbulence in the 2-zone as homogeneous, but
different from that in the 1-zone.

Similarly, one expects that the mean velocity difference across
each next $j$-zone of the width $H_j=H/2^j$ will be more or less
the same. This is the motivation to define ``\ $j$-zone\ " in the
scale invariant manner, as $2^{-j}H<y<2^{-(j-1)}H$  and to
approximate the turbulence inside each such zone as homogeneous.

%Such a scale-invariant subdivision of the channel cross-section is
%reasonable, but somehow arbitrary. We chose here this subdivision
%only for simplicity. The physically motivated choice of the zone
%widths, $h_j$, will appear below, in Sec.~\ref{sss:shell-choice}.

\begin{figure*} %%
\epsfxsize=16.5 cm \epsfbox{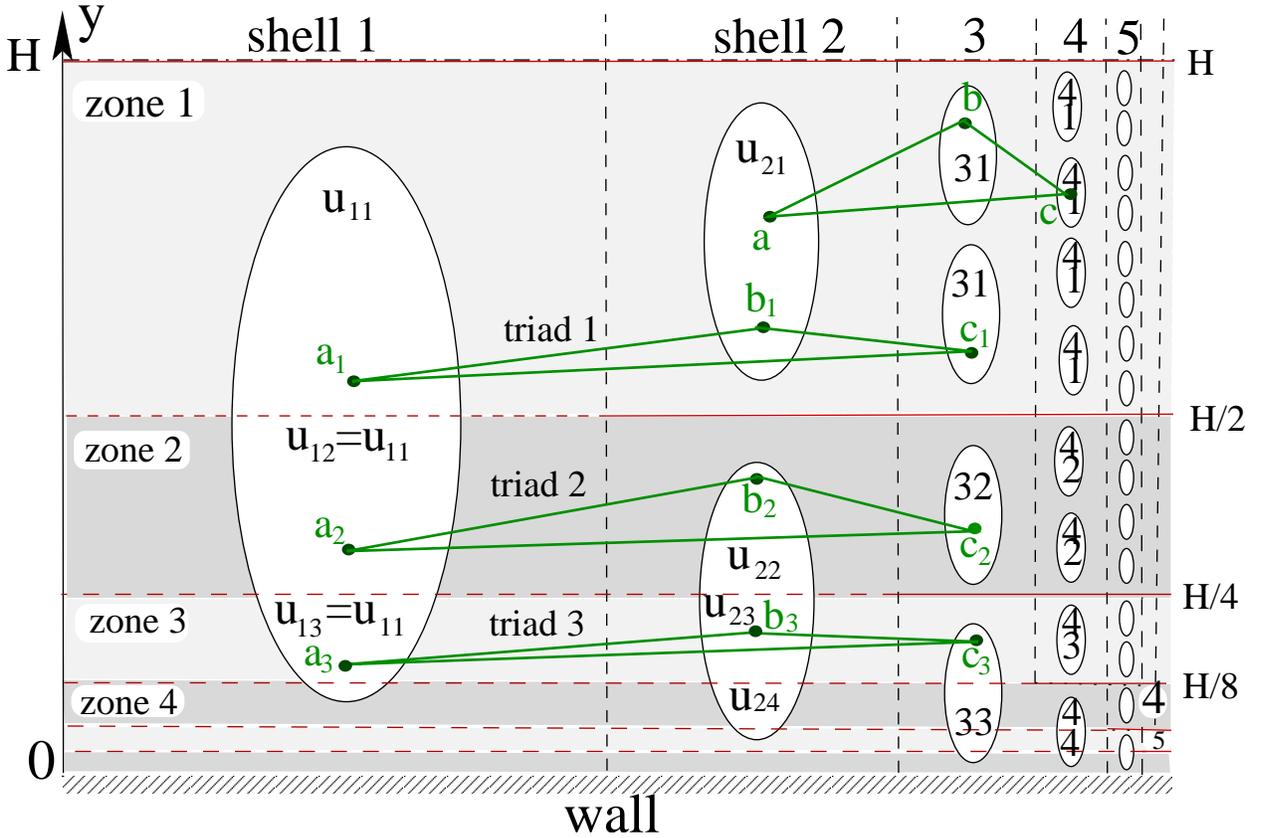} %%
\caption{\label{f:energy-cas} Zones, shells and triad interactions
in the multi-zone shell model for the channel geometry. Regions of
localization of $(n\bm p)$-eddies are shown schematically as ellipses
with corresponding numbers inside. The near-wall eddies have
$n=j$. They also occupy all $j$-zones towards the wall, with
$j>n$.}
\end{figure*}

The approximation of the piecewise homogeneity allows one to use
the shell-model reduction, Eq.~(\ref{eq:simpl-1}), inside of each
$j$-zone, similarly to that in the whole space for homogeneous
turbulence:%%
\bse\label{eq:simpl-2} \bea\label{eq:simpl-2a}%%
U_{n\B p}(t)& \Rightarrow & u_{nj}(t) A_{n\B p_j}\,,\qquad 1\le j
\le n\,,\br
 \overline{A_{n\B p}}&=&0\,, \qquad\overline{|A_{n\B
p}|^2}=1\,, \\ \label{eq:simpl-2b}
 u_{nj}&=&u_{nn}\,, \hskip 2cm  j > n\ .
\EEA \ESE%%
Here $\B p_j$ belong to the $j$-zone,
%$\B p_j\in \C Z_j$,
in the
sense that the  $({n\B p_j})$-cells
 are inside the
$j$-zone. We introduced in \REF{simpl-2a} the velocity of
$(nj)$-eddies,  $n$th shell in the $j$-zone.
Equation~\Ref{simpl-2b} reflects the fact that the near-wall
$(nj)$-eddies with zone index $j>n$ belong simultaneously to the
$j$-zone with $j=n$, see Fig.~\ref{f:energy-cas}. Therefore in our
model all $u_{nj}(t)$ with $j > n$  are just $u_{nn}(t)$.

With Eqs.~\Ref{simpl-2} one gets from Eqs.~(\ref{eq:Pars-b}):
\begin{eqnarray}\label{eq:En} %%%%%%%%
&&  \hskip -1cm \C E_n(t)=\sum_{j=1}^\infty \frac{H_j}H\C E_{nj}(t)\,, \quad
\C E_{nj}(t)=\frac{|u_{nj}(t)|^2}2\,,
\end{eqnarray}
where $\C E_{nj}$ is the energy density of the $n$th shell in the
$j$-zone of the width $H_j=H/2^j$.
%The fact, that \REF{En} for the
%energy in the cell basis has the physically transparent form in
%which $\C E_n$ is un-weighted sum of $\C E_{nj}$ and  $\C E_{nj}$
%is proportional to the zone width $H_j$ is a direct consequence of
%the scale invariant normalization~\Ref{orth-s3}, chosen  for the
%sake of simplicity.  Below, in Sec.~\ref{sss:shell-choice} we
%consider another,  physically motivated bases, which gives
%different, geometry dependent equation for the zone widths.

Using Eqs.~\Ref{simpl-2} and (\ref{eq:shell-eq-exact}) one gets
the equation of motion  of the  multi-zone shell model for
turbulent fluctuations:
\begin{eqnarray}\label{eq:shell-eq-3}
&&  \hskip -1cm \frac{d u_{nj}(t)}{d\,  t }  =  -\gamma _{n}
u_{nj}(t) + \C N_{nj}+\Delta_{nj}w_j\,,
\end{eqnarray}
in which we have added by hand the production term $w_j$,
describing the energy pumping to the turbulent system. This term
will be clarified in the following Section  by \REF{w-n}.

The  nonlinear term $\C N_{nj}$ in \REF{shell-eq-3} describes the
total energy balance for the $n$th shell in the $j$-zone. There
are two distinct geometries: $j<n$ and $j \geq n$. In the latter
case it is enough to describe the energy balance for $j=n$, since
all $u_{nj}=u_{nn}$ for $j>n$.

All eddies with $j<n$ are fully placed  in the same zone (e.g.,
the eddies of shells $n=2$, 3  and 4, that belong to the 1st zone,
see Fig.~\ref{f:energy-cas}) . Therefore in this case we can use
for $\C N_{nj}(t)$ a standard  shell model expression for $N_n$,
in which $u_{nj} \Rightarrow u_n$.  In this paper we adopt the
Sabra version of $N_{nj}$ term, generalizing
Eq.~(\ref{eq:Nn-Sabra}):%%
\bse \label{eq:Nnj-Sabra}\begin{eqnarray} \label{eq:Nnj-Sabra-a}%%
\C N_{nj}(t)&=&N_{nj}(t)\,, \qquad \text{for}\quad  j<n\,,\\
\label{eq:Nnj-Sabra-b}
 N_{nj}&=&i \big[ a \, \kappa _{n+1} u_{n+1,j}^*u_{n+2,j}\br
 && + b \,\kappa _n \,u_{n-1,j}^*u_{n+1,j} -c\,
\kappa _{n-1}u_{n-2,j} u_{n-1,j}  \big]\ .
\end{eqnarray}\ESE%%

The energy balance of the near-wall eddies ($n=j$) is quite
different. As one sees in Fig.~\ref{f:energy-cas} (on the example
of the eddy in the first shell), the near-wall eddies participate
in the triad interactions of three types:\\
-- \emph{Triad~1}~ involves one near-wall eddy and two bulk eddies
[$(u_{11} -u_{21}-u_{31})$ triad of the 1st-zone in the above
example]. For this interactions we will use \REF{Nn-Sabra} but
with different parameters: $a_1$, $b_1$ and $c_1$;\\
-- \emph{Triad~2}~ involves two  near-wall eddies and one bulk
eddy [$(u_{12} -u_{22}-u_{32})$ triad of the 2st-zone in
Fig.~\ref{f:energy-cas}]. Here we will use \REF{Nn-Sabra} with the
parameters: $a_2$, $b_2$ and $c_2$;\\
-- \emph{Triad~3}~ involves three  near-wall eddies
  [$(u_{13} -u_{23}-u_{33})$ triad
of the 3rd-zone in the above example]. Here we will use again
\REF{Nn-Sabra} with the parameters: $a_3$, $b_3$ and $c_3$.

 The relationships between four set of interaction parameters
[($a,b,c$) and $(a_p,b_p,c_p)$, $p=1,2,3$] may be found from: i)
the requirement of the conservation of energy and ii) the
``correspondence principle": \\
\emph{For the space homogeneous case} ($u_{nj}$ is independent of
the zone index $j$) \emph{ the multi-zone model must coincide with
the usual shell model for homogeneous turbulence }(in our case,
with the Sabra model).

The above two requirements give:
%%%%%%
\bea\label{eq:rel-abc} %%
a_1&=&a/2\,, \qquad a_2=a_3=a/4\,, \br %%
b_1&=&b\,, \qquad \quad~ b_2=b_3=b/2\,, \br %%
c&=&c_1=c_2=c_3\ . %%
\EEA %%
These equations may be interpreted as follows:\\
-- in the \emph{Triads 1} only the half of the largest eddy
belongs to the same zone as two smaller ones. This gives
$a_1=a/2$. Two smaller eddies in the \emph{Triad 1} fully belong
to
their zone. This corresponds to $b_1=b$ and $c_1=c$;\\
-- in the \emph{Triads 2} only the quarter  of the largest eddy
belongs to the same zone as two smaller ones. Therefore $a_2=a/4$.
In this triad only the half of the middle eddy belongs to the same
zone as the smallest one.  This corresponds to $b_2=b/2$. The
smallest eddy in the triad fully belongs to their
own zone, and we take $c_2=c$;\\
-- in the \emph{Triads 3} only the  quarter  of the largest eddy,
the half of the middle one and the full smallest eddy belong to
the same set of zones (Zones 3,4,.. in our example). This
corresponds to the relationships $a_3=a/4$, $b_3=b/2$ and $c_3=c$.

The total contribution of three types of the triad interactions to
the nonlinearity of the near-wall eddies can be summarized as
follows:
\begin{eqnarray} \label{eq:Nnn-Sabra}
\C N_{nn}=\frac14\big[ 2 N_{n,n}+ N_{n,n+1}+ N_{n,n+2}\big]\ .&&
\end{eqnarray}
One can join two Eqs.~\Ref{Nnj-Sabra} and  \Ref{Nnn-Sabra} and
write:%%
\bea\label{eq:N-inv}%%
\C N_{nj}= N_{nj}+\Delta_{nj}\Big[\frac14
\big(N_{n,n+1}-N_{nn}\big)~~&&\br + \frac14
\big(N_{n,n+2}-N_{nn}\big)\Big]&& .
\end{eqnarray}%%
Obviously, in the homogeneous case the second term in the RHS of
\REF{N-inv} vanishes and one recovers the usual shell model for
homogeneous turbulence with $\C N_{nj}\Rightarrow N_{n}$.

Notice that the explicit form of this equation reflects the fact
that we have accounted only for the triad interactions involving
the nearest shells ($n-1,\ n$ and $n+1$) and the particular form
of the channel subdivision on the zones: $H_j=2^{-j}H$. This
subdivision  of the cross-section area on zones is reasonable for
the scale invariant case of turbulent boundary layer near flat
plane. The physically motivated subdivision in particular flow
geometries will be discussed in the following section. In general
case \REF{Nnn-Sabra} is changed as follows: %%
\bea\label{eq:N-gen1}%%
\C N_{nj}&=& N_{nj}+\Delta_{nj}\sum_{i=n+1}^\infty
\frac{s_i}{\sigma_j} \big(N_{ni}-N_{nn}\big)\,, \br%%
 \sigma_j&\equiv&\sum_{i=j}^\infty s_i \,,
\end{eqnarray}%%
with an arbitrary dependence of the (dimensionless) $j$-zone
areas, $s_j$, on $j$ and an arbitrary form the nonlinear term
$N_{nj}$.  The fact that under the sum one has the difference
$(N_{n j}-N_{nn})$ guarantees the correspondence principle, while
the weights $s_i/\sigma_j$ follow from the requirement of the
conservation of the total energy, which in the case of arbitrary
zone division is given by a natural generalization of \Ref{En}
with $H_j/H$ replaced by $s_j$:%%
\begin{equation}
 \C E_n(t)=\sum_{j=1}^\infty s_j\C E_{nj}(t)\,,\quad
 \C E_{nj}(t)=\frac{|u_{nj}(t)|^2}2\ .
\end{equation}
\section{\label{s:moment-eq}%%
Dynamical Multi-Scale model for the ``\PM-velocity"}%%

\subsection{\label{ss:Momenta}%%
Mechanical momenta, $\B P$ and $\B M$, \\ and ``statistics vs.
dynamics" dilemma}%%
The unbounded turbulence may be described in the reference system
with a zero mean velocity. In that system the total linear
momentum vanishes: $\B P=0$. Due to the Galilean invariance the
space homogeneous velocity does not interact with the turbulent
fluctuations. Therefore  one can consider turbulence in any other
reference system with $ \B P\ne 0$ with the same result for the
statistic of turbulence. Hence, the mechanical momentum is not a
relevant integral of motion for the unbounded turbulence. This is
not the case for the wall bounded turbulence, in which the
Galilean invariance is broken by the presence of the walls. The
conservation laws for the total linear momentum, $\B P$, (as well
as for the total angular momentum, $\B M$) give an important
constraint on the expected behavior of the system. For example, in
the channel and pipe flows, the total input of the linear momentum
due to the pressure gradient, acting in the whole cross-section
area of the flow, must be equal to the dissipation of the momentum
on the walls due to the viscous friction. Clearly, an analytical
description of the wall bounded turbulence must respect the
conservation of $\B P$ and $\B M$ (in the unforced, inviscid
limit).

Consider the mean velocity field $\B V(\B \rho)$ in the simple
flows with translation symmetry in the streamwise direction $\hat
{\bf x}\| \B P$, in which also $\B M \parallel \hat{\bf x}$. Here
$\B \rho$ is the radius-vector in the cross-section of the flow,
$\B \rho\bot \hat {\bf x}$. The examples of such flows are the
channel and the pipe flows, the planar Couette flow,
Fig.~\ref{f:geometry}, the circular Couette flow, etc. Instead of
$\B P$ and $\B M$, it is more convenient to deal with their volume
densities $\C P$ and $\C M$, defined   as follows:%%
\bse\label{eq:def-ints}\bea\label{eq:def-ints-a}%%
 \C P&=&\C P_x=(\hat{\bf x}, \B V)\,,\\ \label{eq:def-ints-b}
\C M&=&\C M_x=(\BC R, \B V)\,, \quad \BC R \equiv \hat{\bf
x}\times \B \rho \EEA\ESE%%
where $(\B A,\B B)$ is the scalar product~\Ref{def-AB}.

Our idea is to divide the full velocity field $\B V(\B \rho)$ into
two parts, denoted as  $\B V_{_{P\!M}}(\B \rho)$ and
 $\B V\Sb T(\B \rho)$, such that  the ``turbulent part" $\B V\Sb
 T(\B \rho)$ does not contribute to $\C P$ and $\C M$:%%
\bea\label{eq:PM-vel}%%
(\hat{\bf x}, \B V\Sb T)= 0\,, \qquad (\BC R, \B V\Sb T)= 0\,,%%
\EEA%%
 and to take a special care only on  $\B V_{_{P\!M}}(\B
\rho)$ part, contributing to $\C P$ and $\C M$. This division may
be done in many ways, our particular choice will be clarified
below by \REF{expV-PM}.

The description of $\B V_{_{P\!M}}(\B \rho,t)$ may be statistical
or dynamical. The statistical description is the straightforward
``cell-expansion approach to shell models", based on  the NSE for
$\B V_{_{P\!M}}(\B \rho,t)$ in the cell representation, similar to
\REF{shell-eq-exact}. To continue, one has to find some reasonable
statistical simplifications, similar to \REF{simpl-2}.  However,
since of $\C P\ne 0$ and/or $\C M\ne 0$,  now also $\< \B
V_{_{P\!M}}(\B \rho,t)\>\ne 0$  and therefore $\overline{A_{n\bm
p}}$ cannot be approximated as zero. This makes it hardly possible
to remain on the level of statistical description and one has to
deal with the detailed, dynamical description of the
\emph{$P\!M$-velocity} $\B V_{_{P\!M}}(\B \rho,t)$  in terms of
the NSE. This is the subject of the following Subsections.

\subsection{\label{ss:PM-basis}%%
\PM-basis, properties and interpretation}%%

\subsubsection{\label{sss:def-bas}%%
Construction  of  the \PM-basis}%%
To construct a ``physically motivated" basis for description of
the \PM-velocities, consider two subsets of eigenfunctions of the
Laplace  operator, $\B \phi^+_m(\bm\rho)$ and $\B
\phi^-_m(\bm\rho)$, satisfying the incompressibility condition.
The no-slip boundary conditions are assumed in the cross-section
of
the flow, with the constraints%%%%
\bse\label{eq:cont}\bea\label{eq:cont-a}  %%
p_m&=&(\hat { \bf  x },\B \phi^+_m)\ne 0\,, \\
\label{eq:cont-b}%%
\C R_m&=& (\BC R, \B \phi^-_m)\ne 0 \,, %%
\EEA\ESE%%
dictated by \REF{PM-vel}. Notice that both  sets $\B \phi^+_m$ and
$\B \phi^-_m$ are chosen here as real and orthonormal.

Introduce the complex $P\!M$-basis as follows: %%
\bse\label{eq:Phi-n} \bea\label{eq:Phi-n-a} %%
\B \Phi_j(\bm\rho) &=& \sum_{m\in\mathcal
S_j}\left[p_m\bm\phi^+_m -i\frac{\C R_m}{R_j}{\bm\phi}^-_m
\right] \,,\\ \label{eq:Phi-n-b}%%
 m  \in \C S_j &\Leftrightarrow&
2^{j-1} \le m \le 2^j-1 \ .%%
\EEA\ESE%%
 The normalization ``radius" $R_j$, is chosen such that%%
\begin{eqnarray}\label{eq:ort-1}%%
(\B \Phi_j',\B \Phi_j')&=&(\B \Phi_j'',\B \Phi_j'')\,, \EEA %%
where%%
\[
 \B \Phi_j' (\bm\rho)\equiv\text{Re}[\B
\Phi_j(\bm\rho)]\,, \quad \B \Phi_j'' (\bm\rho)\equiv\text{Im}[\B
\Phi_j(\bm\rho)]\ .%%
\]
Denote %%
\bea\label{eq:den:hj}%%
s_j\equiv (\B \Phi_j',\B \Phi_j')=\sum_{m\in \C S_j} p_{m}^2\ .%%
\EEA%%
We show below, that%%
\bea\label{eq:s-sum}%%
\sum_{j=1}^\infty s_j=1\,, %%
\EEA%%
that allows us to understand \emph{$s_j$ as a portion of the
cross-section area, occupied by the $j$-zone}.  Then, the first of
Eqs.~\Ref{def-s} gives the definition of $S_j$, the area of
$j$-zone, that is consistent with the requirement
$\sum_{j=1}^\infty S_j=S_\bot$.

To find $R_j$ from \REF{ort-1} compute also %%
\bea\label{eq:den:hj-1}%%
(\B \Phi_j'',\B \Phi_j'')=\frac1{R_j^2}\sum_{m\in \C S_j} \C
R_{m}^2\ . %%
\EEA%%
Then %%
\begin{eqnarray}%%
\label{eq:h-n1-b}%%
R_j^2&=& s_j^{-1}\sum_{m\in \C S_j}\C R_{m}^2  \ .
\end{eqnarray}%%
\subsubsection{\label{sss:ortho}Orthogonality and normalization conditions}
By construction~(\ref{eq:Phi-n}--\ref{eq:ort-1}) the $P\!M$-basis
is orthogonal in the sense:
\begin{subequations}\label{eq:ortho}
\begin{eqnarray}
 (\bm\Phi_j,\bm\Phi_{j'}) &=& 2 s_j\Delta_{jj'}
\,,\\
 (\bm\Phi_j^*,\bm\Phi_{j'}) &=& 0 \ .\end{eqnarray}
\end{subequations}
The idea behind  the choice~\Ref{Phi-n}  is that the functions $\B
\Phi_j'$ and $\B\Phi_j''$ form the \emph{exact}  expansions of the
uniform profile of unit height ($\B\Phi_j'$) and of the linear
profile ($\B\Phi_j''$):%%
\begin{subequations}\label{eq:unit}
\begin{eqnarray}\label{eq:unit-a}
 \sum_j \bm\Phi_j'(\bm\rho) &=& \hat{\bf  x}
\,,\\ \label{eq:unit-b}
 \sum_jR_j \bm\Phi_j''(\bm\rho) &=& -\BC R\ .%%
\end{eqnarray} \end{subequations}%%
These relations can be easily proven by projection of both sides
of \REF{unit} onto the Laplace basis $\bm\phi^\pm_m$,
using \Ref{cont} and the definition of the
$P\!M$-functions $\bm\Phi_j(\bm\rho)$, \REF{Phi-n}.

 Let us show that \REF{s-sum} follows from the obvious constrain
 $(\x,\x)=1$, after substitution  $\x$ from \REF{unit-a}. Using
 also the orthogonality conditions~\Ref{ortho} one gets%%
\begin{eqnarray*}
1&=&(\x,\x)=\sum_{j,j'=1}^\infty (\B \Phi_j',\B \Phi_{j'}') \br %%
&=&\frac14 \sum_{j,j'=1}^\infty (\B \Phi_j+\Phi_j^*,\B
\Phi_{j'}+\B \Phi_{j'}^*)= \sum_{j=1}^\infty s_j\ . %%
\end{eqnarray*}
We proved that all $s_j$ add up to unity according to \REF{s-sum}.

\subsubsection{\label{sss:PM-exp}\PM-expansions}
Introduce the \PM-projector%%
\bea\label{eq:PM-proj}%%
\hat {\B P}\SPM  \{\B f (\B \rho)\}&=&   2\, \text{Re} \sum_j
\frac{(\B \Phi_j, \B f) } {( \B \Phi_j, \B \Phi_j)} \B \Phi_j(\B
\rho)\br%%
&=&\text{Re} \sum_j s_j^{-1}(\B \Phi_j, \B f)\, \B \Phi_j(\B \rho)
\, \,,~~~~~~~~~~~ %%
 \EEA
and define the  $P\!M$-velocity $\B V_{_{P\!M}}(\B \rho,t)$ as the
projection of the full field $\B V(\B \rho,t)$ on the
$P\!M$-basis:%%
\bea\label{eq:V-PM} %%
 \B V\SPM (\B \rho) =\hat {\B P}\SPM
 \{\B V(\B \rho)\}\ .%%
 \EEA%%
Using the normalization~\Ref{ortho} one gets %%
\bse\label{eq:expV-PM} \bea\label{eq:expV-PM-a} %%
\B V_{_{P\!M}}(\B \rho,t)&\equiv &\text{Re}\Big\{\sum_{j=1}
^\infty V_{j}(t) {\B \Phi}_{j}(\B \rho)\Big\}\,,\\
\label{eq:expV-PM-b} %%
s_j\,V_{j}(t)&\equiv&(\B \Phi_j, \B V)=(\B \Phi_j,\B V_{_{P\!M}})\
.
\end{eqnarray}\ESE%%
The expansion coefficients $V_j(t)$ can be understood as \PM
velocities in the zone representation. They play the major role in
our theory and will be called hereafter ``the \PM-velocities" or
``zone velocities".

Substituting   $\hat{\bf  x}$ and $\BC R$ from Eqs.~\Ref{unit}
into the definitions~\Ref{def-ints} of $\C P$ and $\C M$,  one
gets with the help of Eqs.~\Ref{expV-PM} the \PM-expansion of the
linear in $\B V$ integrals of motion:%%
\begin{subequations}\label{eq:ints-1}\begin{eqnarray}\label{eq:P-1}%%
\C P &=& \sum_{j=1}^\infty
 ( \B\Phi'_{j},\B V)= \sum_j  s_j \,V_{j}'(t)
\,,\\ \label{eq:M-1}%%
\C M &=&  \sum_{j=1}^\infty R_j
 (\B\Phi''_{j},\B V)=\sum_j  s_j \,  (R_j V_{j}''(t))\ .%%
 \end{eqnarray}\end{subequations}%%
This is a proof of the statement that  only $P\!M$ part of the
full velocity $\B V(\B \rho,t)$, \REF{expV-PM}, contribute to the
mechanical momenta. The turbulent part $\B V\Sb T(\B \rho)$\emph{
does not contribute } to the linear integrals of motion and will
be considered as a part of the ``turbulent ensemble", described by
the shell variables $u_{nj}(t)$.

The energy density associated with the $P\!M$-velocity is as
follows:%%
\bea\label{eq:E-PM}%%
&&\hskip -1.65cm{\C E}_{_{P\!M}}(t)\equiv \int_{S_\bot} \frac{|\B
V_{_{P\!M}}(\B \rho,t)|^2}2  \frac {d\B \rho}{S_\perp}=
\sum_{j=1}^\infty %%
   s_j \frac{|V_j(t)|^2}2\ .%%
\end{eqnarray}%%

\subsubsection{\label{sss:plane-pipe-bas}%%
\PM-basis for the channel and pipe flows}%%
The particular form of the \PM-functions is geometry dependent. To
give a feeling how these functions may look like, we discuss here
two important cases, the channel and the pipe flows. These two
examples will serve us in the rest of the paper. Some properties
of these specific bases are however more general and will be used
in the derivation and analysis of the momentum equation for
$V_j(t)$.

In the planar geometry, Fig.~\ref{f:geometry}, $\B \rho
\Rightarrow y$ and the functions $\B \phi^\pm_m$ in \REF{Phi-n} are given by%%
\bse\label{eq:F-bas}\bea\label{eq:F-bas-a} %%
\B \phi^+_m(y)&=&\hat{\bf x}\phi_{2m-1}(y)\,,\br %%
\B \phi^-_m(y)&=&\hat{\bf z}\phi_{2m}(y)
 \,,\br \phi_m(y)&=& \sqrt 2 \sin (k_m
y)\,,
\quad k_m={\pi\, m}/{(2 H)}\,,\\
\label{eq:F-bas-b}%%
p_m&= & \frac{ 2 \sqrt 2 }{\pi\, (2 m-1)} \simeq 0.9 /(2m-1)\ .%%
\EEA\ESE%%
For the pipe geometry, %%
\bse\label{eq:F-pipe}\bea\label{eq:F-pipe-a}%%
\B \phi^+_m(\rho)&=&\hat{\bf x}\phi_{m}(\rho)\,, \quad %%
\B \phi^-_m(\rho)=\hat{\bf e}\phi_{m}(\rho)\,, \\
\label{eq:F-pipe-b}%%
\phi_m(\rho) &=&  {J_0(k_m\rho )}\Big /{J_1(k_m R)}\,,
 \\ \label{eq:F-pipe-c}%%
 J_0(k_m R)
&=&0\,,\quad~~~~~~~~~ p_m =  2/(k_m R)\,,%%
\EEA\ESE%%
where $\hat{\bf e}$ is the polar-angle unit vector ($\hat{\bf
e}\perp \hat{\bf x} $, $\hat{\bf e} \perp \hat{\B \rho}$),
$J_0(\xi)$ and $J_1(\xi)$ are the Bessel functions of the 0-th and
1-st order, and $R$ is the radius of the pipe. First of
\REF{F-pipe-c} defines $k_m=\xi_m/R$ via zeros of the Bessel
functions $\xi_m$ : $J_0(\xi_m)=0$.%%
\begin{figure*} %%
\epsfxsize=8.6cm \epsfbox{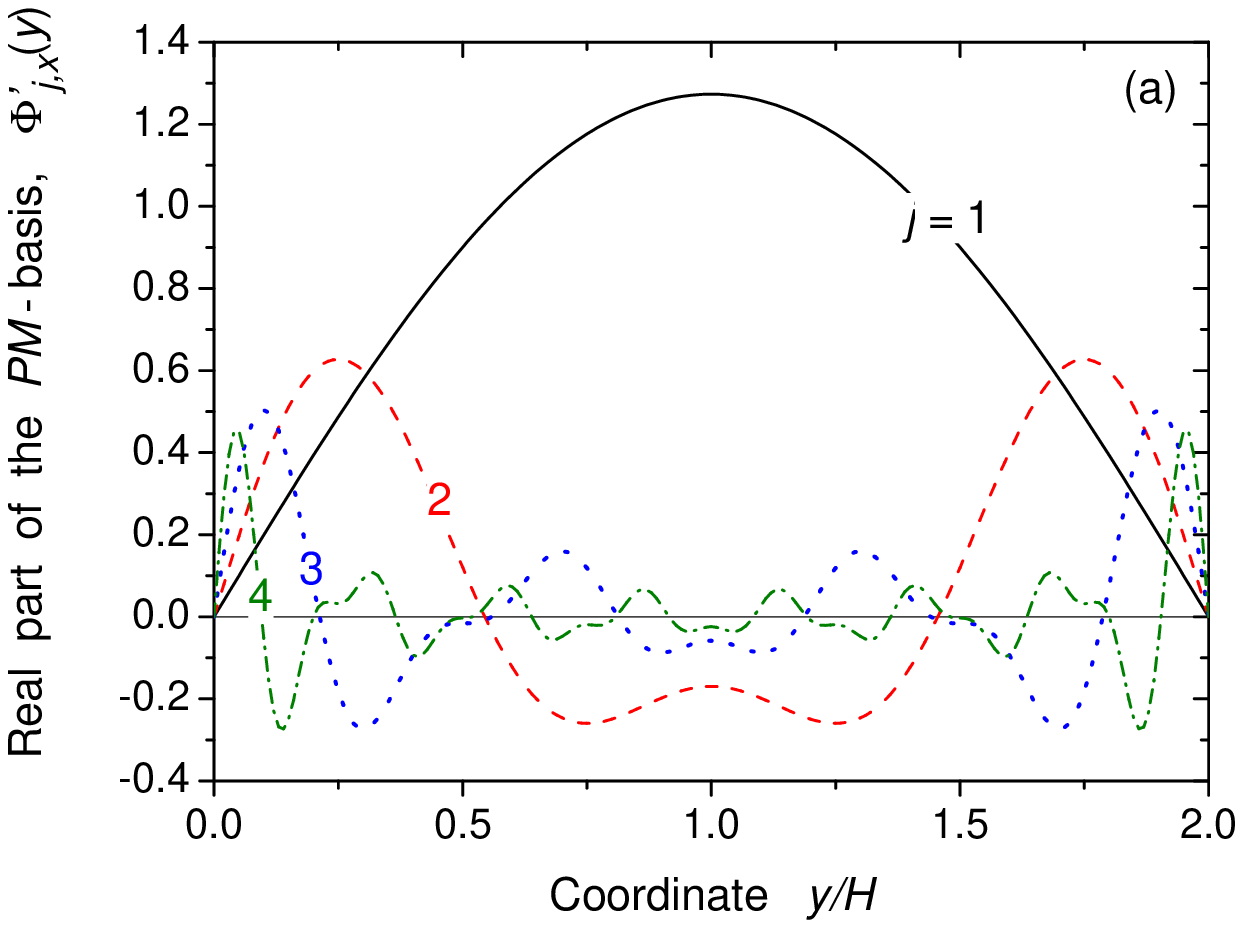} %%
\epsfxsize=8.6cm \epsfbox{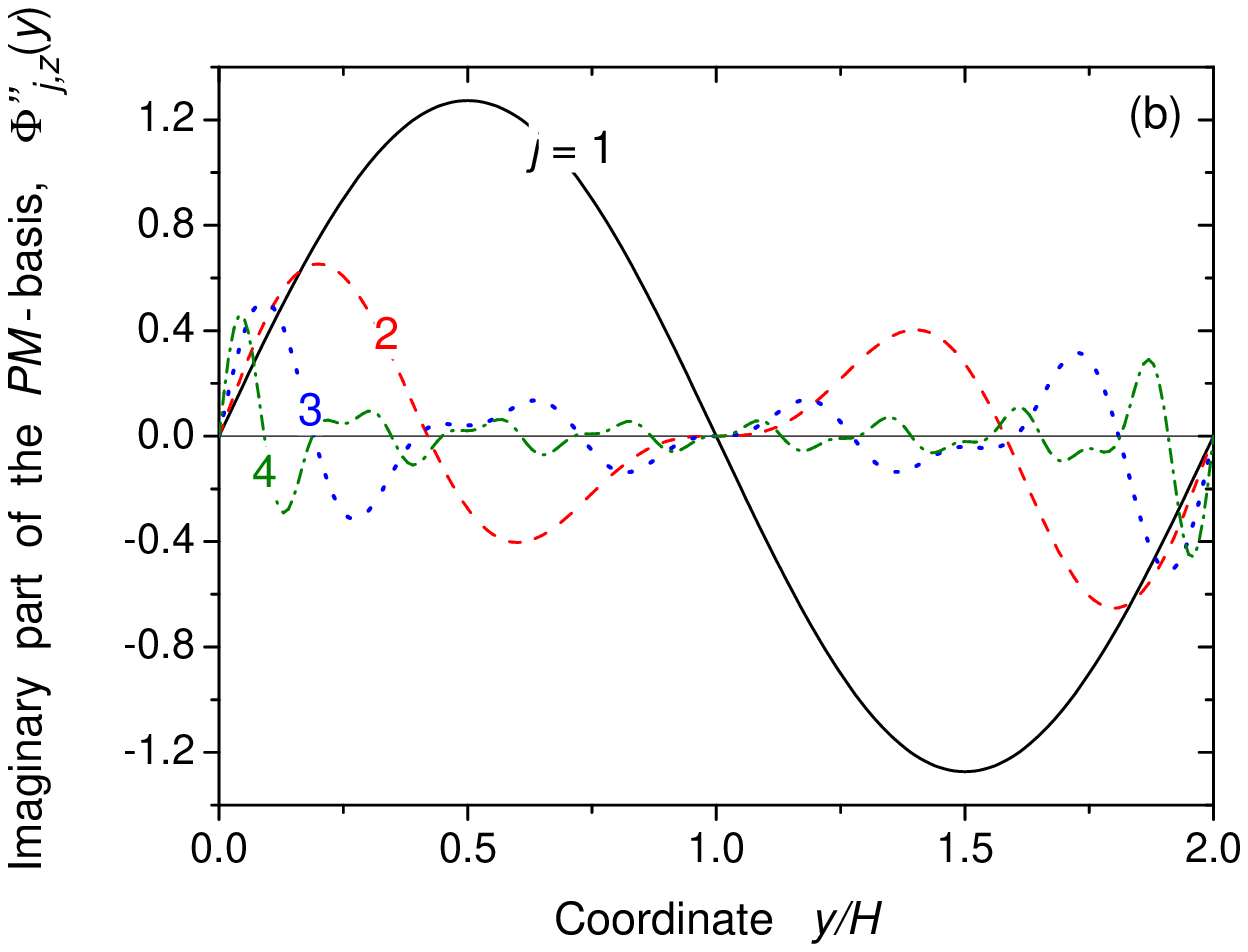} %%
\epsfxsize=8.6cm \epsfbox{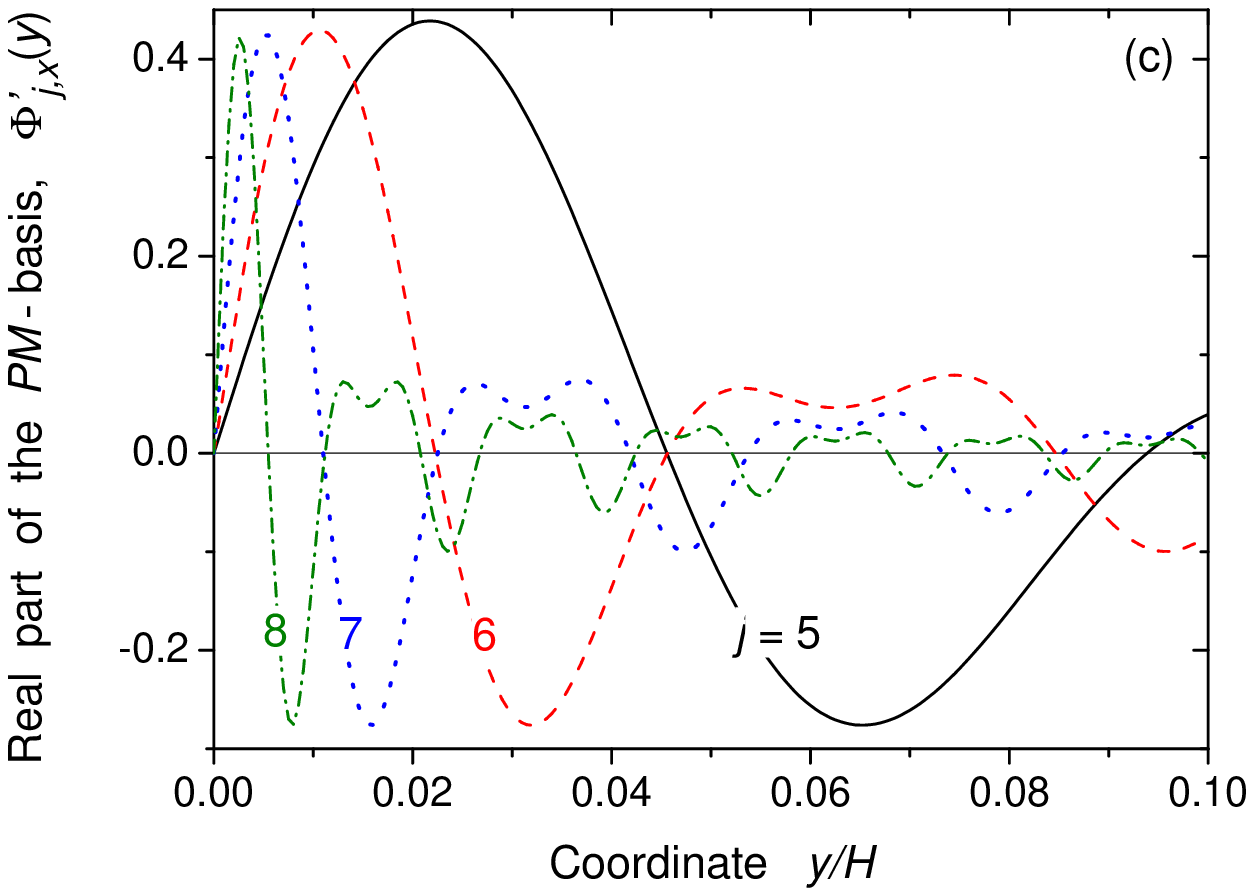} %%
\epsfxsize=8.6cm \epsfbox{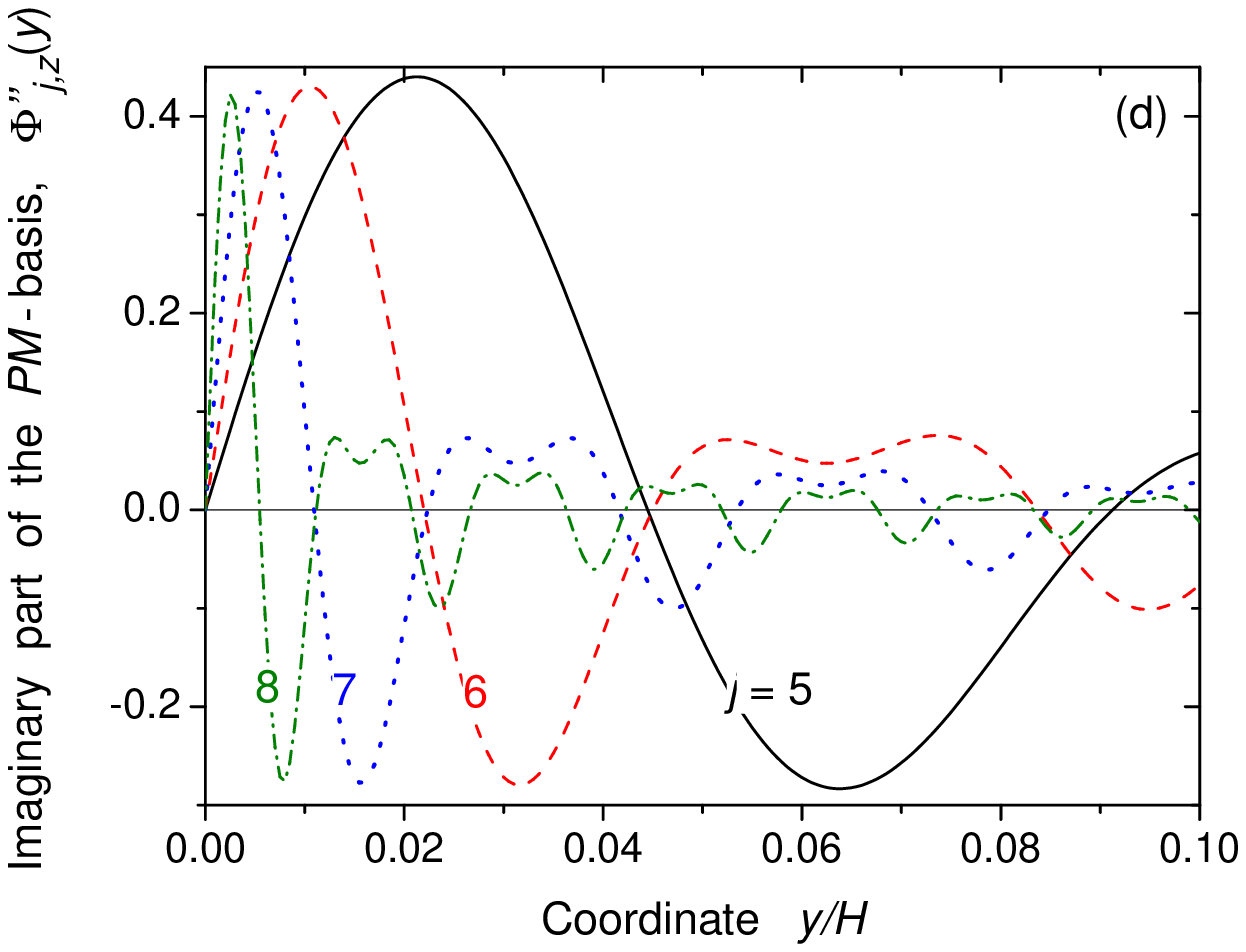} %%
\caption{\label{f:new-basis}%%
First functions $\B \Phi_{j}(\xi H)$ describing the mean flow with
non-zero mechanical momenta in the channel of the width $2\,H$. In
the whole channel $0<\xi< 2$. Panels $a$ and $b$: the
\PM-functions
 $\Phi_{j,x}'(\xi H)$   (Panel $a$) and $\Phi_{j,z}''(\xi H)$
 (Panel $b$) with $j=1,2,3,4$ for the whole channel. Panels $c$
 and $d$: the \PM-functions  with
$j=5,6,7,8$ in the near-wall region. Panel $c$: $\Phi_{j,x}'(\xi
H)$.  Panel~$d$: $\Phi_{j,z}''(\xi H)$.}
\end{figure*}%%

In Fig. \ref{f:new-basis} we plot $\Phi_{j,x}'(y)$ and
$\Phi_{j,z}''(y)$ for the channel as it follows from \REF{Phi-n}
with \REF{F-bas}. The functions $\Phi_{j,x}'(y)$ are symmetric
with respect to the centerline  of the channel $y=H$, whereas
$\Phi_{j,z}''(y)$ are antisymmetric. For large $j$,
$\Phi_{j,x}'(y)$ and  $\Phi_{j,z}''(y)$ coincide  in the lower
half of the channel $y<H$ and have an opposite sign for $y>H$.  As
it is clear from Panels $c$ and $d$, in the near-wall ($y\ll H$)
region $\Phi_{j,x}'(y)$ and $\Phi_{j,z}''(y)$  almost coincide
already for $j=5$.%%
\begin{figure*}%%
 \epsfxsize=8.6cm \epsfbox{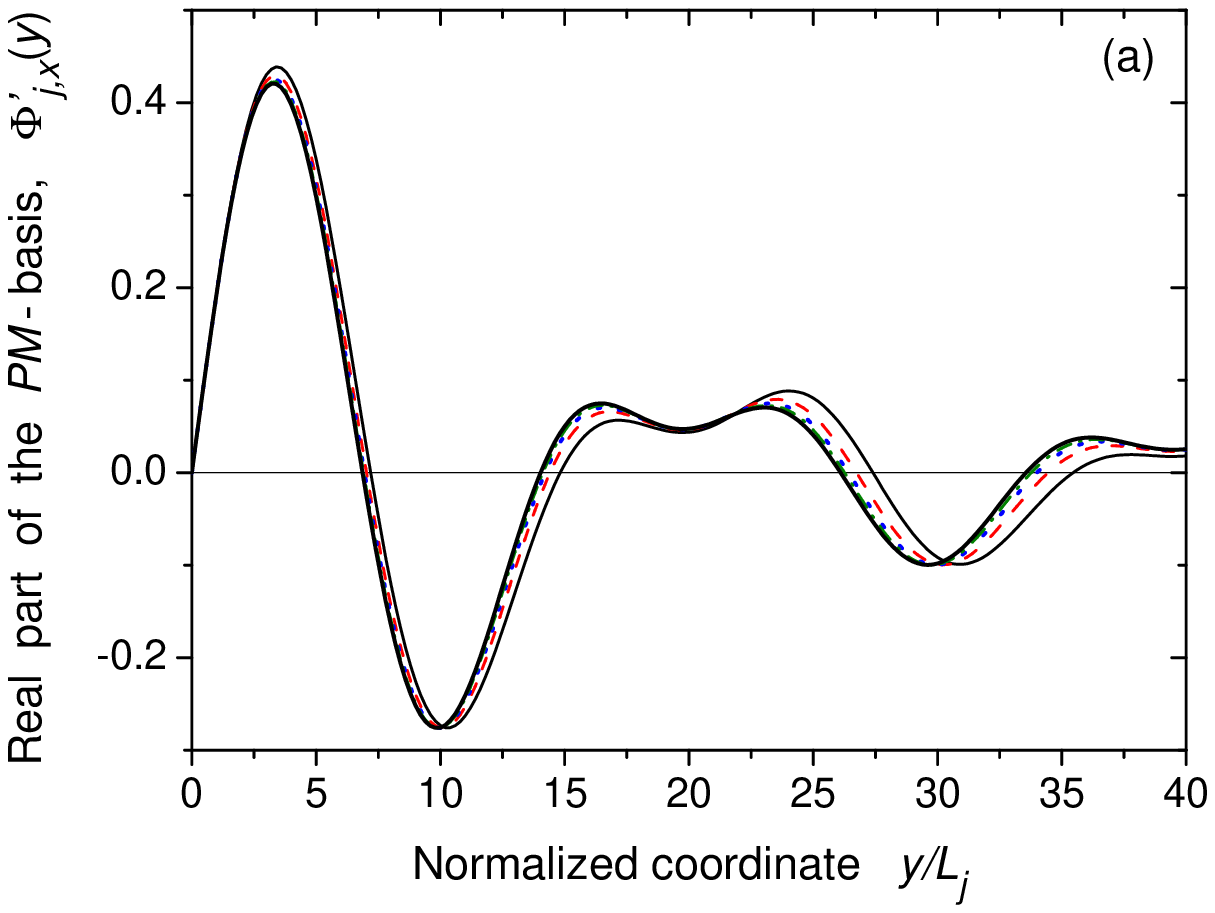}%%
  \epsfxsize=8.6cm\epsfbox{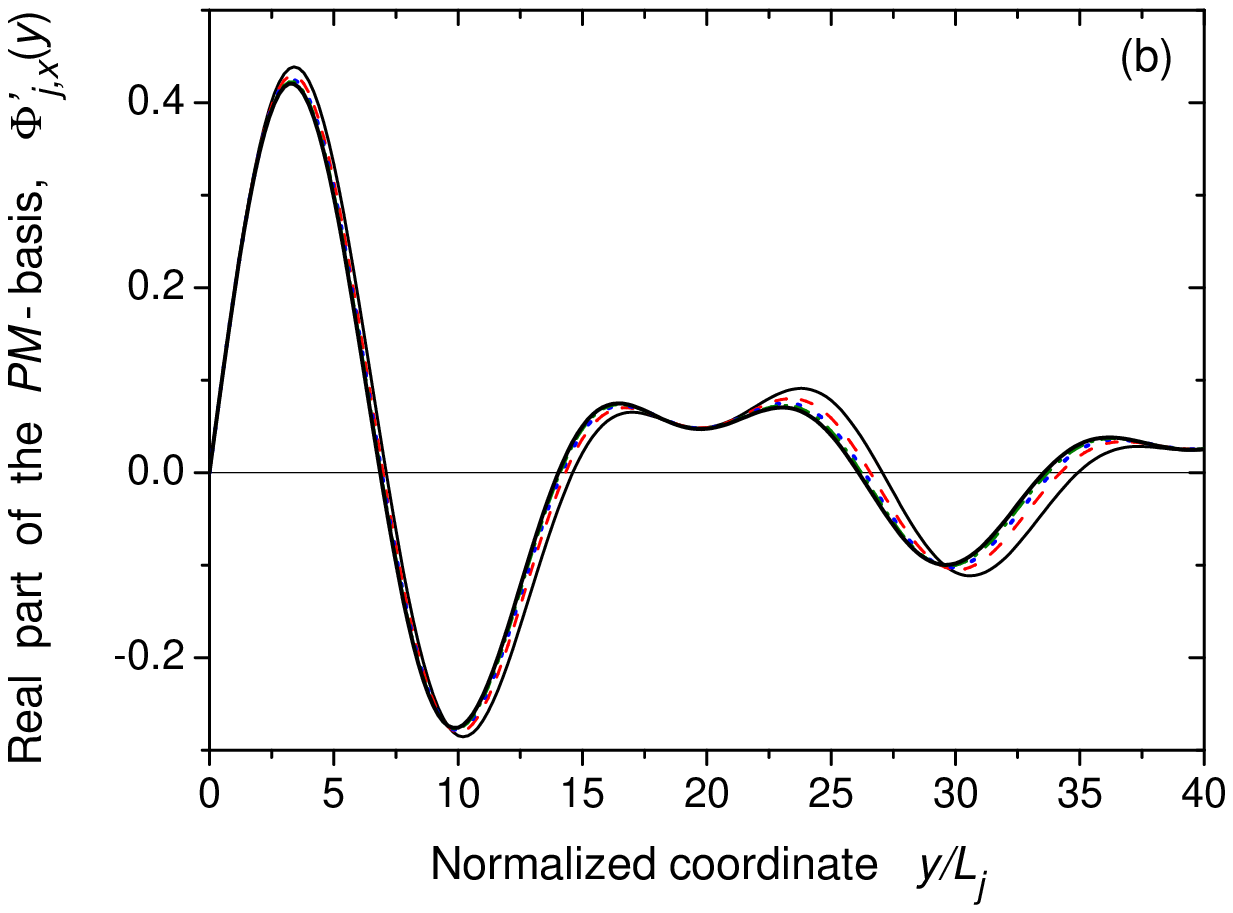} %%
\caption{\label{f:new-collapse}%%
Asymptotic universality of the \PM-basis functions.  Panel $a$:
Collapse of the re-scaled  real part of the basis functions
$\Phi_{j}  (2H\xi/\pi^22^j )$ for $n=5...8$ with the universal
asymptotic function $\Phi_{\text{un}}(\xi)$ for the channel  flow.
The lines (from right to left) correspond to those in
Fig.~\ref{f:new-basis}c. The left-most solid line denote
$\Phi_{\text{un}}(\xi)$.  Panel $b$: the same for the pipe flow,
$H=R$. }
\end{figure*}%%
For the channel flow  one can find an explicit asymptotic
expression for $\Phi_{j,x}'(y)$ and $\Phi_{j,z}''(y)$ in the limit
$j\to \infty$ (actually, $j>4$). For $y<H$:
\begin{eqnarray}\label{eq:as-F}%%
\Phi_{j,x}'(y)&\simeq& \Phi_{j,z}''(y)
  \simeq \Phi_{\rm un}(2^j\pi^2y/2H)
\,,\\\label{eq:univ-b}
 \Phi_{\rm un}(\xi) &\equiv&
   \frac2\pi\left[\text{Si}\left(\frac{2\xi}{\pi}\right)
    -\text{Si}\left(\frac{\xi}{\pi}\right)\right]
\,,\end{eqnarray}%%
where $\text{Si}(x)$ is the sine integral function.

It is expected, that the asymptotic form of basic functions,
$\Phi_{\rm un}$, is the same for any flow geometry, if one express
them as functions of the distance from the wall. For example, in
the pipe one obtains \Ref{as-F}, where $y$ is the distance from
the wall, $y=R-\rho$, and $H$ is the radius of the pipe, $H=R$.
In Fig. \ref{f:new-collapse} we show the collapse of the re-scaled
functions $\Phi_j(2H\xi/\pi^22^j)$ for $j=5,\dots 8$ with the
universal function $\Phi_{\text{un}}(\xi)$ for the channel (Panel
$a$) and pipe (Panel $b$, $H=R$) flows.%%

\subsubsection{\label{sss:zones}%%
Geometry of the $j$-zones in the channel and pipe flows}

\paragraph{Characteristic length of a flow.}
The cross-section of the flow can be characterized by two global
parameters, the cross-section area, $S_\bot$, and the length of
its perimeter $P_\bot$.  One can organize from these two objects
many combinations with the dimensions of the length, $L(\beta)
\equiv P_\bot (S_\bot/ P_\bot ^2 )^\beta$, with arbitrary $\beta$.
For discussion of a turbulent flows under the external pressure
gradient, the particular choice, $\beta=1$, %%
\bea\label{eq:L}%%
L\equiv L(1)=S_\bot/ P_\bot\,,%%
\EEA%%
is physically important. The reason is that the total external
accelerating force, applied on the unite length of the fluid (in
the streamwise direction), is proportional to $ S_\bot$: %%
\[
F\sb{ac}= \nabla p S_\bot\,,
\]
while the total friction force is proportional to $ P_\bot$ (under
simplifying assumption of homogeneity along the perimeter):%%
\[
F\sb{fric}= \nu_0 P_\bot \frac{d V_x(y)}{dy}\Big|_{y=0}\ .%%
\]
The stationarity condition, $F\sb{ac}= F\sb{fric}$, allows one to
relate $[d V_x(y)/dy]_{y=0}$ to the characteristic length, $L$
defined by \REF{L}:%%
\[%%
\frac{d V_x(y)}{dy}\Big|_{y=0}=\frac{\nabla p}{\nu_0\, L}\ .%%
\]
In terms of the friction velocity, $U_\tau$, \REF{def-TV}, and the
friction Reynolds number, $\RE$, \REF{def-RE}, this gives the
famous constraint  for the wall bounded flows:%%
\bea\label{eq:con}%%
\frac{d V_x(y)}{dy}\Big|_{y=0}= \RE \frac{U_\tau}{L}\ . %%
\EEA%%
In all mentioned global relationships the characteristic length
$L$ plays an important role.

Notice that for the channel of the width $2H$ the length $L$ is
 the distance from the centerline to a wall, $L=H$. For the pipe of
the radius $R$ the length $L=R/2$,  which is twice smaller than
the distance from the center to the wall.%%
\paragraph{\label{p:zones}%%
Zones in the channel and pipe.} %%
In Tab.~\ref{t:coefficients} we present parameters $s_j$ for the
channel and pipe geometries, that are given by \REF{den:hj} via
$p_m$. The parameters $p_m$ are taken from \REF{F-bas-b} for the
channel, and from \REF{F-pipe-c} for pipe geometry.%%
\begin{table}
\caption{\label{t:coefficients} Parameters of the \PM-basis for
the planar   and pipe  geometries.  }
\begin{tabular}{||c||c|c|c|c||c|c|c|c||}
\hline\hline
 &\multicolumn{4}{|c||}{Planar}&\multicolumn{4}{c||}{Pipe}
 \\\cline{2-9}
 \raisebox{1.5ex}[0pt]{~$j$~} &
 ~$  s_j$~ & $~s_j2^j~$ & $R_j/H$ & $\nu_j/\nu_0$ &
 ~$  s_j$~ & $~s_j2^j~$ & $R_j/R$ & $\nu_j/\nu_0$ \\
\hline
 1 & ~0.811~ & ~1.62~ & ~0.50~ & ~6.48~ &
     ~0.692~ & ~1.38~ & ~0.51~ & ~2.77~ \\
 2 &  0.122  &  0.49  &  0.77  &  1.96  &
      0.185  &  0.74  &  1.03  &  1.48  \\
 3 &  0.038  &  0.30  &  0.90  &  1.22  &
      0.068  &  0.54  &  1.00  &  1.09  \\
 4 &  0.015  &  0.25  &  0.95  &  0.99  &
      0.029  &  0.47  &  1.00  &  0.94  \\
 5 &  0.007  &  0.22  &  0.98  &  0.89  &
      0.014  &  0.44  &  1.00  &  0.87  \\
 6 &  0.003  &  0.21  &  0.99  &  0.85  &
      0.007  &  0.42  &  1.00  &  0.84  \\
 7 &  0.002  &  0.21  &  0.99  &  0.83  &
      0.003  &  0.41  &  1.00  &  0.82  \\
 8 &  0.001  &  0.21  &  1.00  &  0.82  &
      0.002  &  0.41  &  1.00  &  0.82  \\
\hline $\infty$
   & -- & $2/\pi^2$ & 1 & $8/\pi^2$
   & -- & $4/\pi^2$ & 1 & $8/\pi^2$ \\
\hline\hline
\end{tabular}
\end{table}%%

Using $s_j$ and $L$ we introduce the \emph{characteristic width of
the $j$-zone}:%%
 \bea\label{eq:Lj}%%
L_j\equiv s_j L\ .%%
\EEA%%
The idea behind this definition is that for  the narrow near-wall
zones $L_j$ is exactly their geometrical width, $\Delta_j$, i.e.
the distance between the $j$-zone boundaries.

Assuming that $\Delta_j$ is much smaller than  the local curvature
of the boundary, we evaluate the area of very narrow $j$-zone as
$S_j= \Delta_j\,P_\bot$. On the other hand, this area is $s_j
S_\bot$. Therefore,%%
\[\Delta_j=s_j S_\bot /P_\bot=s_j\,L= L_j\,,\qquad \text{for}\
 j\gg 1\ .%%
\]
Obviously, for the planar geometry, $\Delta_j=L_j$ for any $j$. To
find the value of $j$, at which  the zones become
flat, consider the pipe geometry with $\Delta_j=r_j-r_{j-1}$,
where $r_j$ is the radius of the circle, occupied by the first
$j$-zones: %%
\bea\label{eq:bound} %%
r_j&=&R\Big[\sum_{i=1}^j s_i\Big]^{1/2}\,, \quad \text{pipe} \ .
\EEA%%
Table~\ref{t:DR} compares $\Delta_j/R$, given by this equation
with $L_j/R= s_j/2$. Clearly,  for $j>3$ these two parameters
practically coincide.

\begin{table}
\caption{\label{t:DR} Geometrical and characteristic widths
$\Delta_j$ and $L_j$ of $j$-zone in the pipe geometry}
\begin{tabular}{||c||c|c|c|c|c|c||}\hline\hline
 $j$&1&2&3&4&5&6\\ \hline\hline
 $\Delta_j/R$&~~0.83~~&~~0.105~~&~0.0356~&~0.0148~&~0.0071~&~0.0035~\\ \hline
$L_j/R$&~0.35 &~0.093~& ~0.0340& ~0.0145&~0.0070&~0.0035\\
\hline\hline
\end{tabular}
\end{table}

\subsubsection{\label{sss:inter}%%
Interpretation of the \PM-velocities}

The \PM-expansions of the linear profiles~\Ref{unit}  gives a
simple interpretation of the velocities $V_j'(t)$ and $V_j''(t)$
as the $\hat{\bf x}$-  and $\hat{\bf e}$-projections of $\B V\SPM
(\B \rho,t)$ at some position within the $j$-zone, $\rho_j$:%%
\bse \label{eq:inter} \bea\label{eq:inter-a} %%
&& \hskip -2cm V_j'(t)\Leftrightarrow V_x(\rho_j,t)\,,\quad
V_x(\B
\rho,t)\equiv \hat{\bf x}\cdot \B V\SPM(\B \rho,t)\,, \\
\label{eq:inter-b}%%
&& \hskip -2.05cm  V_j''(t)\Leftrightarrow  V\sb e(\rho_j,t)\
,\quad V\sb e(\B
\rho,t)\equiv \hat{\bf e} \cdot \B V\SPM(\B \rho,t)\ , \\
\label{eq:inter-c}%%
&& ~~~~~~~~~~~~~~~~~~~ \hat{\bf e}  \equiv \hat{\bf x}\times \B
\rho/\rho \ .%%
\EEA\ESE%%
This point will be illustrated below in
Fig.~\ref{f:Vy-eps-y-profs} by direct comparison of $V_j$ and
$V(\B \rho)$ in appropriate coordinates.

\subsection{\label{ss:simple-Vj-eq}%%
Simple momentum equation for $V_j$} %%
Consider the NSE for $\B V\SPM (\bm\rho,t)$ in the form:%%
\bea\label{eq:NSE} %%
\frac{\partial \B V\SPM (\B \rho,t)}{\partial t} &=&-\nu_0 \Delta
\B V\SPM
 +\nabla p \, \hat {\bf x}\br &&- (\B V\SPM \cdot \B \nabla)\B V\SPM
- \overline{(\B u \cdot \B \nabla)\B u}\ .%%
 \EEA%%
Here $\nu_0$ is the kinematic viscosity,  $\nabla p=-dp/dx>0$ is the pressure
gradient in the streamwise direction $\hat {\bf x}$. The nonlinear
term $(\B V\SPM \cdot \B \nabla)\B V\SPM $  describes the
self-nonlinearity in the \PM-subsystem and the last term,
$\overline{(\B u \cdot \B \nabla)\B u}$, is responsible  for the
effect of the Reynolds stress on the \PM-velocity. Here
$\overline{\cdots}$ is the ensemble averaging in the sense of
\REF{simpl-2}. The nonlinear cross term, $\propto V u$ (which has
zero mean) is neglected.

The goal is to get the equation of motion for the complex
\PM-velocities $V_j(t)$:%%
\begin{eqnarray}\label{eq:res-1}%%
\frac{dV_j}{dt} &=&~\text {(damping)}_j+ \text{(pressure)}_j\br &&
+\text{(self-interaction)}_j+\text{(Reynolds-stress)}_j \,,%%
\end{eqnarray}%%
by projecting   NSE~\Ref{NSE} on the ${P\!M}$-basis, using
Eqs.~\Ref{ortho} and \Ref{expV-PM-b} in the form:%%
\bea\label{eq:proj} %%
V_j(t)=2(\B \Phi_j,\B V\SPM)/(\B \Phi_j,\B \Phi_j)\ .%%
\EEA%%

\subsubsection{\label{sss:demping}The damping term}%%
Projecting the NSE~\Ref{NSE} on the real and imaginary parts of
the \PM-basis~\REF{Phi-n}, and accounting only for the viscous
term, $\propto \nu_0$, one gets%%
\[%%
\frac{dV_j'}{dt} = -\Gamma_j' V_j'\,, \quad \frac{dV_j''}{dt} =
-\Gamma_j'' V_j'\,,%%
\]%%
with, generally speaking, different $\Gamma_j'$ and
$\Gamma_j''$:%%
\bea\label{eq:res-2}%%
 \Gamma_j'= -\nu_0\frac{\left(\bm\Phi_j',\Delta\bm\Phi_j'\right)}
 {\left(\bm\Phi_j',\bm\Phi_j'\right)}\,,\ \Gamma_j''=
 -\nu_0\frac{\left(\bm\Phi_j'',\Delta\bm\Phi_j''\right)}
 {\left(\bm\Phi_j'',\bm\Phi_j''\right)}\ .%%
\EEA%% %%
However, in the pipe flow they are equal, $\Gamma_j'=\Gamma_j''$,
since the functions $\B \phi^+_m$ and $\B \phi^-_m$ are the same,
see \REF{F-pipe}. In the channel flow these functions,
\REF{F-bas}, are different and hence $\Gamma_j'\ne\Gamma_j''$.
Nevertheless, for large $j$, the basis functions $\Phi_{j}'\to
\Phi_{j}''$ and $\Gamma_j'\to \Gamma_j''$. Therefore it would be a
reasonable simplification to neglect the (possible) difference
between $\Gamma_j'$ and $\Gamma_j''$ and to write in
\REF{res-1}:%%
\bea\label{eq:dam} %%
\text {(damping)}_j= -\Gamma_jV_j\,, %%
\EEA%%
with the same damping term $ \Gamma_j$, which is in between
$\Gamma'_j$ and $\Gamma''_j$. For example, in the channel flow,
$V_j'$ is responsible for the mean velocity profile $V_x(y)$ and
thus ``more important" than  $V_j''$. In that case we will take
$\Gamma_j=\Gamma'_j$. In the planar Couette flow, the mean
velocity profile is given by $V_z(y)$, which is connected to
$V_j''$, and thus one better take $\Gamma_j=\Gamma''_j$.

It is customary to represent the damping term via an effective
viscosity, $\nu_j$, and an effective wave-vector, $\kappa_j$,
defined via the characteristic width of the $j$-zone $L_j$: %%
\bea\label{eq:nu-j} %%
\Gamma_j&=&\nu_j \kappa_j^2\,, \qquad \nu_j \sim \nu_0 \,,
\\ \label{eq:kappa}%%
\kappa_j&\equiv &\frac{1}{2\, L_j}\,, \quad L_j\equiv s_j\,L \ .%%
\EEA %%
Parameters $L$ and  $L_j$ were  defined and discussed in
Sec.~\ref{p:zones}. The values of $s_j$ and $\nu_j$ for the
channel and pipe flows are presented in Tab.~\ref{t:coefficients}.

Notice, that the damping term in \REF{dam} is diagonal in $j$.
This is a consequence of our definition of the \PM-functions
$\bm\Phi_j$,  \REF{Phi-n}: the functions with \emph{different} $j$
originate from different, and thus \emph{orthogonal}
eigenfunctions of the Laplace operator.%%
\subsubsection{\label{sss:pressure}%%
The pressure term}%%
The equations \Ref{NSE} and \Ref{proj} dictate: %%
\[%%
\text{(pressure)}_j=2 \nabla p \,(\B \Phi_j,\hat{\bf x})/(\B \Phi_j,\B
\Phi_j)\ .%%
\]%%
Substituting  $\hat{\bf x}$ from \REF{unit-a} into  above
expression, one gets for any flow geometry the same simple
expression: %%
\bea\label{eq:res-3}%%
\text{(pressure)}_j= \nabla p\ .%%
\EEA %%
This means, that the uniform in  $\B \rho$ pressure gradient
equally acts on all components $V_j'(t)$. This remarkable result
could be also obtained from the interpretation of $V_j$,
Sec.~\ref{sss:inter}. Indeed, if we leave in the RHS of \REF{NSE}
only the pressure term, we get the $\B \rho$-independent  $V_x(\B
\rho,t)=\nabla p\, t$, i.e.  a homogeneous velocity profile.
This means that  $V'_j(t)$ is independent of $j$.%
\subsubsection{\label{sss:self-inter}%%
The self-interaction term}%%
In various simple flows  the self-interaction term in \REF{NSE} is
identically equal to zero due to geometrical constraint. It is so
for the channel and the planar Couette flows, where %%
\[(\B V\cdot \B \nabla)\B V\Rightarrow
\left(V_x\frac{\partial}{\partial x}+V_z\frac{\partial}{\partial
z}\right)\B V(y)\equiv 0\ .
\]
Here  we skipped for the brevity the subscript ``$\SPM$".  The
same is true for the pipe geometry, where
\[(\B V\cdot \B \nabla)\B V\Rightarrow
\left(V_x\frac{\partial}{\partial x}+\frac{V_\phi}{ \rho}
\frac{\partial}{\partial \phi} \right)\B V(\rho)\equiv 0\ .
\]
In the present paper we consider only flows with zero
self-interaction: \bea\label{eq:res-4}%%
\text{(self-ineraction)}_j= 0\ .%%
\EEA %%
\subsubsection{\label{sss:interaction}%%
Effect of the Reynolds stress}%%
In the \PM-representation the Reynolds-stress term in \REF{res-1}
should be quadratic function of the turbulent velocities $u_{nj}$
and should scale as $\kappa_j\equiv 1 /(2L_j)\propto2^{j}$. This
term describes two physical processes, namely, (i) the exchange of
the linear and angular momenta between different zones and, (ii)
together with its counterpart $w_j$ in the equation for $u_{nj}$,
the energy exchange between $V_j$- and $u_{nj}$-subsystems.

In the spirit of the shell models of turbulence,  we consider here
the form the Reynolds-stress term that accounts for the momenta
exchange only between nearest $j$-zones ( $j$ and $j\pm1$) and
preserves the relevant integrals of motions, the energy and the
momenta. Notice, that  in the turbulent channel and pipe flows
with $\< \C M \>=0$, the conservation of $\C P$ is much more
important that the conservation of $\C M$. Therefore we can
simplify the possible structure of the inter-system interaction
terms, $W_j$ and $w_j$ by using instead of \REF{M-1} for $\C M$
its simplified version, in which all $R_j$ are assumed to be the
same, $R_j\to R_\infty={\rm const}$, and thus can be omitted:%%
\bea\label{eq:quasi-M}%%
&&\hskip -1.3cm \tilde {\C M} = \sum_j s_j V_j'' \ \Rightarrow \
\text{Quasi-angular momentum}. %%
\EEA%%
It is easily justified by noting  that for $j>3$ the $j$-zone
contributions to $\C M$ and $\tilde{\C M}$ are almost the same,
see Table~\ref{t:coefficients} for $R_j$. In the following we will
drop the tilde in $\~{\C M}$,
 and write quasi-angular
momentum simply as $\C M$.
 Having in mind the
conservation of the ``complex momentum":
\[
\C L\equiv \mathcal P +i\,{\mathcal M} = \sum_j s_jV_j\,,
\]
we suggest the following form of the Reynolds-stress
term, $W_j$,:
\bea\label{eq:res-5}%%
\text{(Reynolds-stress)}_j\equiv W_j=\frac1{s_j}\left(\F
p_{j-1}-\F p_j\right) \,,%%
\EEA %%
 where $\F p_{j}$ is the momentum flux from the  $j$- to the
$(j+1)$-zone.  Indeed, this form
 provides the conservation of $\C L$  in the inviscid,
unforced limit:%%
\[
\frac{d\C L}{dt} = \sum_j s_jW_j = \sum_j(\F p_{j-1}-\F p_j) = 0\
.%%
\]%%

For the momentum flux $\F p_j$ we also select the simplest form,
assuming that only the near-wall turbulent eddies of the same
scale $u_{jj}\equiv u_j$ (i.e. $u_{nj}$ with  $n=j$) give a
nonzero momentum transfer:%%
\bea\label{eq:Jn}%%
\F p_j = \frac{d}{2 L} u_j^2 \ .%%
\EEA%%
Here the dimensionless parameter $d$ characterize the  strength
of the interactions.
% They are evaluated in Sec.~\ref{s:analysis}
%by considering the problem of stability of the laminar flow.

The equations \Ref{res-5} and (\ref{eq:Jn}) give us the following
simple expression for $W_j$: %%
\bea\label{eq:W-n}%%
 W_j =d\,\kappa_j(u_{j-1}^2-u_{j}^2) \ .%%
 \EEA

The production term $w_j$, describing  the energy pumping into
$j$-zone of the turbulent subsystem, \REF{shell-eq-3}, is then
uniquely determined by the requirement of the conservation of the
total energy of the system:%%
\bea\label{eq:w-n}%%
w_j = \frac{d}{2 \sigma_jL}(V_j-V_{j+1})\, u_j^* \ .%%
\EEA%%%%

\subsubsection{\label{sss:mom-eq}%%
Resulting momentum equation}%%
Collecting Eqs.~\Ref{res-1}, \Ref{dam} -- \Ref{res-4} and
\Ref{res-5} one finally gets:%%
\bea\label{eq:mom-j}%%
\frac{d V_j(t)}{d t} &\!\!\!=\!\!\!& -\nu_j \kappa_j^2 V_n +\nabla p +
W_j\,, %%
\EEA%%
where $\nu_j$ is the effective viscosity, given for the channel
and the pipe flows in Table~\ref{t:coefficients}, $\kappa_j$ is
given by \REF{kappa} and $W_j$ -- by \REF{W-n}. This equation
together with \REF{shell-eq-3} will be analytically and
numerically analyzed in the following Secs.~\ref{s:analysis} and
\ref{s:sol}.

%% SECTION III

\section{\label{s:analysis}%%
 Conservation laws and symmetries in the MZS-model }%%

%%%%%%%%%%%%%%%%%%%%%%%%%%%%%%%%%%%%%%%%%%%%%
\subsection{\label{ss:res}%%
Resulting dimensionless MZS-equations}
For the convenience of the reader we present here the full set of
the MZS-equations which will be  the subject of analytical and
numerically studies.  As the first step,  we non-dimensionalize
the MZS-equations, expressing   time $t$ and velocities $V_j$,
$u_{nj}$ in  the  units  $\tau$ and $U_\tau$:
\begin{subequations}\label{eq:norm} \begin{eqnarray}\label{eq:norm-b}%%
\tilde t ~~ &\equiv& ~~~~ \frac{t}{\tau} \,,  \\
\label{eq:norm-c}%%
\tilde V_j (\tilde t ) &\equiv&  \frac {V_j (\tau\,\tilde t) }{U_\tau}
\,,\\
\tilde u_{nj} (\tilde t ) &\equiv&  \frac {u_{nj} (\tau\,\tilde t)
}{U_\tau} \,,\\ \label{eq:norm-d}%%
\nabla\tilde p&=&1\,,
\end{eqnarray}
\end{subequations}
where  time, $\tau$, and velocity, $U_\tau$,  are constructed from
the ``outer" characteristics of the flow, $\nabla p$ and $L$:%%
\bea\label{eq:units} \tau  &\equiv& \sqrt{L/\nabla p}\,,  \qquad U_\tau
\equiv  \sqrt{\nabla p\, L}\ . %%
\EEA%%
In the rest of the paper we omit the tildes in the dimensionless
variables.  The  dimensionless MZS-equations take on the form:
\begin{subequations}\label{eq:MSM}
\begin{eqnarray}\label{eq:MSM-a}
  \frac{dV_j}{dt} &=& -\Gamma_jV_j +\nabla p_j +W_j
\,,\quad \nabla p_j=\nabla p=1\,, \\ \label{eq:MSM-b}
  \frac{du_{nj}}{dt} &=& -\gamma_nu_{nj} +\mathcal N_{nj}
  +\Delta_{nj}w_j \,,\br %%
j&=&1,2,\dots \infty\,,\quad n=j,j+1,\dots \infty\ .
\end{eqnarray}
\end{subequations}
Here the  shell variables $u_{nj}$ are  the velocities of
statistically identical ${(nj)}$-eddies of characteristic scale
$s_n$ that belong to $j$-zone of width $s_j$. Clearly, in
Eqs.~\Ref{MSM-b} $n\ge j$. The \PM-variables $V_j(t)$ describe the
velocities of coherent near-wall structures in the $j$-zone. In
our notations the near wall turbulent $(nj)$-eddies have $n=j$.
These eddies  occupy also all $j$-zones with $j\ge n$. Therefore
in our approach
\[%%
u_{nj}=u_{nn}\equiv u_n\,,\quad\text{for}\qquad j\ge n\ .
\]
The terms in the RHS of the dimensionless MZS-equations~\Ref{MSM}
are as follows:
\begin{subequations}\label{eq:nots}
\begin{eqnarray}\label{eq:nots-a}
  \Gamma_j &=& \frac{G_j}{\RE}\kappa_j^2
\,,\quad
  \gamma_n = \frac{g_n}{\RE}\kappa_n^2
\,,\quad \kappa_j\equiv \frac1{2 s_j}\,, \\ \label{eq:nots-b}
  W_j &=& d\,\kappa_j\left(u_{j-1}^2-u_j^2\right)
\,,\\ \label{eq:nots-c}
  w_j &=& \frac{d}{2\sigma_j}\left(V_j-V_{j+1}\right)u_j^*
\,,\quad   \sigma_j \equiv  \sum_{j'\ge j}s_{j'}\,, \\
\label{eq:nots-d}
  \mathcal N_{nj} &=& N_{nj}
    +\Delta_{nj}\sum_{j'>j}\frac{s_{j'}}{\sigma_j}\left(N_{nj'}-N_{nj}\right)
\,,\\ \label{eq:nots-ae} N_{nj} &=&
i\big(\,a\,\kappa_{n+1}u_{n+1,j}^*u_{n+2,j}
\\\nonumber&&
    +b\,\kappa_nu_{n-1,j}^*u_{n+1,j}
    -c\,\kappa_{n-1}u_{n-2,j}u_{n-1,j}\big)\ .
\end{eqnarray}
\end{subequations}
Here
\[a+b+c=0\,,\quad \sum_{j\ge1}s_j =1\ .%%
\]%%
The equations~\Ref{nots} contain only one physical parameter, the
friction Reynolds number%%
\bea\label{eq:Re} %%
\RE\equiv L U_\tau /\nu_0\,,%%
\EEA%%
and a  set of geometry-dependent, dimensionless factors $G_j$,
$g_n$, and $s_j$. For the channel and pipe flows the factors
$G_j=\nu_j/\nu_0$ and $s_j$ are given in
Table~\ref{t:coefficients}. For simplicity in this paper we take
 $g_n=G_n$.%%
\subsection{\label{s:fluxes}%%
Conservation laws and fluxes}%%
In the invisid ($\Gamma_j=\gamma_n=0$), unforced ($\nabla p_j=0$)
limit \REF{MSM} conserves energy, $\C E$, linear momentum, $\C P$,
and (quasi-) angular momentum, ${\C M}$:%%
 \bse\label{eq:mot-ints}%%
\bea\label{eq:mot-ints-a}%%
\C E &=& \frac12 \sum_{j=1}^\infty  s_j\Big( |V_j|^2
+\sum_{n=1}^\infty|u_{nj}|^2\Big)\,,%%
\\ \label{eq:mot-ints-b}%%
\C P &=& \sum_{j=1}^\infty s_j V_j' \,,
\\ \label{eq:mot-ints-d}%%
{\C M}& = &\sum_{j=1}^\infty s_j V_j'' \ .%%
\EEA\ESE%%
In general case, with non-zero $\Gamma_j$, $\gamma_j$, and $\nabla
p_j$ the direct calculation of  $d \C E/dt$, $d \C P/dt$, and $d
{\C M}/dt$ with the help of \REF{MSM} gives:%%
\bse\label{eq:t-ints}%%
\bea\label{eq:t-ints-a}%%
\frac{d \C E}{dt} &=& \ve^+-\ve^-\,,  \\ \label{eq:t-ints-b}%%
\frac{d \C P}{dt} &=& \F p^+-\F p^-\,,  \\ \label{eq:t-ints-c}%%
\frac{d {\C M}}{dt} &=& {\F m}^+-{\F m}^-\ . %%
\EEA\ESE%%
Here $\ve^+$, $\F p^+$, and ${\F m}^+$, are \emph{total} influxes
of respective  integrals of motion, $\C E$, $\C P$, and ${\C M}$,
and $\ve^-$, $\F p^-$, and ${\F m}^-$, are their \emph{total}
rates of dissipations. Similar to Eqs. \Ref{mot-ints} these
objects can be represented as $s_j$-weighted sums,
$\sum_js_j(\dots)$, of the respective densities of \emph{partial}
influxes ($\ve^+_{j}$, $\F p^+_{j}$, and ${\F m}^+_{j}$)  and
\emph{partial} rates of dissipation in the $j$-zone:%%
\bse\label{eq:p-inf} \bea \label{eq:p-inf-a} %%
\varepsilon^+ &=&\sum_{j=1}^\infty s_j \, \varepsilon^+_{j}\,,
\qquad \varepsilon^+_{j}=\nabla p_j\,V_j'= V_j' \,,
\\ \label{eq:p-inf-b}%%
\F p^+ &=&\sum_{j=1}^\infty s_j \, \F p^+_{j}\,, \qquad \F
p^+_{j}=\nabla p_j=1\,,%%
\\ \label{eq:p-inf-c}%%
{\F m}^+ &=&\sum_{j=1}^\infty s_j \, {\F m}^+_{j}\,, \qquad
{\F m}^+_{j}=0\,,%%
\EEA \ESE %%
\bse\label{eq:p-inf-2} \bea \nonumber %%
\varepsilon^- &=&\sum_{j=1}^\infty s_j \,
\varepsilon^-_{j}\,,\qquad \varepsilon^-_{j}=\Gamma_j\,|V_j|^2 +
\sum_{n=1}^\infty \gamma_n \,|u_{nj}|^2\,,\\ \label{eq:p-inf-d}
\\ \label{eq:p-inf-e}%%
\F p^- &=&\sum_{j=1}^\infty s_j \, \F p^-_{j} \,, \qquad \F
p^-_{j}=\Gamma_j V_j'\,,%%
\\ \label{eq:p-inf-f}%%
{\F m}^- &=&\sum_{j=1}^\infty s_j \, {\F m}^-_{j}\,, \qquad
{\F m}^-_{j}=\Gamma_j V_j''\ .%%
\EEA \ESE %%
The total influx  of energy is exactly equal to the total linear
momentum of the flow (due to our normalization, $\nabla p=1$) and
the total influx of the linear momentum equals to unity: %%
\bse\label{eq:influx}\bea\label{eq:influx-a}%%
\varepsilon^+ &=& \nabla p \sum_{j=1}^\infty s_j V_j' = \nabla p\, \C P= \C
P\,,%%
\\ \label{eq:influx-b}%%
\F p^+ &=& \nabla p\sum_{j=1}^\infty s_j~~~= ~\nabla p ~~=1\ . \EEA\ESE%%
For the channel and pipe flows, the influx of the angular
momentum, $\F m^+=0$. In the planar Couette flow, $\F p^+=0$ and
$\F m^+\ne 0$.

Notice, that the main part of the influx of the momentum is always
flowing  into the first few zones, where $s_j$ have considerable
values. For example, in the channel, $\approx 81\%$ of the total
$\F p^+$ flows into the first zone (occupying $s_1\approx 0.81$
part of the cross-section area, see Table~\ref{t:coefficients}),
93\% -- into the first two zones and only about 1\% to the fifth
and all higher zones. As we show below, $V_j'$ slightly decay with
$j$, hence the influx of energy is even more confined to the first
zones.

For very high Reynolds numbers  the dissipation of the conserved
quantities  occurs at large $j$ (or $n$), $\sim \log_2\RE$.
Therefore for $\log_2\RE\gg 1$  there exist an inertial interval
(generally speaking, different for the different quantities).
Clearly, in the stationary case the influx of some conserved
quantity have to be equal to its flux (in the shell space or via
zones).  For the energy, it is useful to analyze the energy flux
in the shell space: %%
\bea\label{eq:stat}%%
\ve^+=\ve_n=\ve^-=\C P\,, \quad \ve_n=\sum_{j=1}^\infty
\ve_{nj}\,,%%
\EEA %%
where $\ve_{nj}$ is usual for the shell models (in the $j$-zone)
expression for the flux of energy from $n$th to $(n+1)$ shell.

The flux of the linear momentum, $\F p_j$ from $j$- to
$(j+1)$-zone is given by \REF{Jn}, which in the dimensionless form
reads: $\F p_j=d\,u_j^2/2$. Therefore, in the stationary case, for
$j$ in the inertial interval %%
\bea \label{eq:const-p}%%
\F p_j&=& d\,u_j^2/2 =\F p^+=\F p^-=1\ . %%
\EEA%%
To get the expression for the  value of $\F p_j$, valid for any
$j$ (not only in the inertial interval), we multiply
  \REF{MSM-a} by the $j$-zone
area, $s_j$, and sum up  from 1  to $j$. This gives the useful
equation %%
\bea\label{eq:p-in}%%
\F p_j=d\,u_j^2/2 = \sum_{i=1}^js_i\left(1-\Gamma_i V_i\right) \,,
\quad \text{any}\ j\,,\EEA%%
that allows us to express  the turbulent velocity $u_j=u_{jj}$,
generating the turbulent energy cascade in the $j$-zone,  via
\PM-velocities.
\subsection{\label{ss:Galilei}%%
Galilean invariance of the model}%%
The Galilean transformation to the reference system, moving with
some velocity $U_0$ in the streamwise direction $\x$, changes $\B
V(\B \rho,t)\Rightarrow \Sb G\B V(\B \rho,t)+ \x U_0$. According
to Eqs.~\Ref{unit-a} and \Ref{proj} it leads to the
transformation: %%
\bea\label{eq:Gal-t1} %%
V_j(t)\Rightarrow  \Sb GV_j(t)+U_0\ .%%
\EEA%%
Clearly, the term $w_j$, \REF{nots-c}, describing the effect of
the mean velocity on the turbulent subsystem, is Galilean
invariant:%%
\bea\label{eq:Gal-t2}%%
 w_j &=& \frac{d}{2\sigma_j}\left(V_j-V_{j+1}\right)
 u_j^*\Rightarrow \Sb G w_j \ .
\EEA%%
In other  words, the uniform velocity profile,  $\B V(\B
\rho)=\x\times $const,  does not affect the statistics of
turbulence. This important invariance of the MZS-equations
guarantees   that for $\RE\to \infty$ the mean velocity profile in
the core of the flow becomes uniform and the statistics of
turbulence becomes space homogeneous, as expected.

\subsection{\label{ss:as-univ}%%
Asymptotic scale invariance and equations for the Turbulent
Boundary Layer near flat plane}%%
Consider MZS-Eqs.~\Ref{MSM} at very large $\RE$ in the near-wall
region, $n,j> j_*\gg 1$, in which the dimensionless parameters in
\REF{nots} can be already replaced by their asymptotic values:%%
\bea \label{eq:as-con} %%
G_j& \Rightarrow & G\,, \qquad~ g_j \Rightarrow g\,,\br  %%
s_j & \Rightarrow & 2^{-j}s\,, \quad \sigma_j=2 s_j\ . %%
\EEA%%%%
This is definitely so for the channel and pipe flows, say, for
$j_* = 3$, see  Tab.~\ref{t:coefficients}. For large $j$ the zone
width is much smaller than the local curvature radius of the wall
and the discussed situation corresponds to the case of {\em
Turbulent Boundary Layer} (TBL) near the flat plane.

For $j\ge j_*$ the dimensionless flux of linear momentum from $j$-
to $(j+1)$-zone,  $\F p_j$, is very close to 1 and much larger
than the direct influx into $j$ zone $s_j\F p^+_{j}=s_j\ll 1$ from
the external pressure gradient. Therefore in this regime one can
neglect in the RHS of \REF{MSM-a} $\nabla p=1$ with respect to
$W_j$ and  simplify \REF{MSM} to the scale-invariant form:
\begin{subequations}\label{eq:iMSM}
\begin{eqnarray}\label{eq:iMSM-a}
  \frac{dV_j}{dt} &=& -\frac{G \kappa_j ^2 }{\RE}V_j
+d\,\kappa_j (u_{j-1}^2 -u_j^2) \,,\\ \label{eq:iMSM-b}
  \frac{du_{nj}}{dt} &=& -\frac{g\kappa_n^2 }{\RE} u_{nj}
  + N_{nj}\br %%
&& \hskip -1.5cm +\frac{\Delta_{nj}}2  \, \Big[\, d\,\kappa_j
(V_j-V_{j+1}) \,
u^*_j + \sum_{j'>j} 2^{j-j'}
\big(N _{nj'}-N_{nj}\big )\Big] \,,\\
\label{eq:iMSM-c}%%
\kappa_j&=&2^j\, \kappa  \,,\qquad \kappa= 1/2 s\,, \quad n\ge
j\ge j_*\,,%%
\\ \label{eq:iMSM-d}%%
 \quad \F p^+&=& d\,u_{j_*-2}^2/2=  d\,
u_{j_*-1}^2/2 =1\,,
\end{eqnarray} \end{subequations}
where $N_{nj}$  has already the scale invariant
form~\REF{nots-ae}.  The ``TBL boundary condition" at
$j=j_*$,~\Ref{iMSM-d},   provides  the influx of energy and
mechanical momentum into TBL,~\REF{iMSM}. In initial \REF{MSM}
this role is played  by the external pressure gradient $\nabla
p=1$.

Equations \Ref{iMSM} have additional [with respect to \REF{MSM}]
rescaling symmetry, namely, they remain unchanged under the
transformation
\begin{eqnarray}\label{univ-sym}
  V_j &\to& \~V_j = V_{j+j_0}
\,,\\\nonumber
  u_{nj} &\to& \~u_{nj} = u_{n+j_0,j+j_0}
\,,\\\nonumber
  t &\to& \~t = t/2^{j_0}
\,,\\\nonumber
  \RE &\to& \~{\RE} = \RE/2^{j_0}
\,,\end{eqnarray} %%
which corresponds (in the $\bm r$-space) to the simultaneous
rescaling of the outer scale $L$ and the pressure gradient $\nabla
p$ in a way that leaves the value of the wall shear stress
unchanged, $\nabla p\,L= \text{const}$. This symmetry of the
equations means that
 MZS-model    describes the asymptotic universality of the near-wall
turbulence for $\RE\to\infty$, and that the only relevant
parameter in this regime is the total influx of the momentum, $\F
p^+$, which is fixed by the boundary conditions \Ref{iMSM-d}.

In studies of  the near-wall turbulence, it is customary to
normalize the time and velocity units by the ``inner'' scales
instead of the ``outer'' ones. Namely, we should use the {\em
viscous length-scale} $\delta\equiv\nu_0/U_\tau=L/\RE$ instead of
the outer scale $L$, and the corresponding timescale
$\tau_\delta\equiv\delta/U_\tau=\tau/\RE$ instead of $\tau$. As
one can see, this rescaling corresponds to the choice
$j_0=\log_2\RE$ in the transformation (\ref{univ-sym}). In new
units, the MZS-equations have the same form as \Ref{iMSM} with
$\RE=1$:
\begin{subequations}\label{eq:rMSM}
\begin{eqnarray}\label{eq:rMSM-a}
  \frac{dV_j}{dt} &=& - G \kappa_j ^2 V_j
+d\,\kappa_j (u_{j-1}^2 -u_j^2) \,,\\
\label{eq:rMSM-b}
  \frac{du_{nj}}{dt} &=& - g\kappa_n^2  u_{nj}
+ N_{nj}\br %%
&& \hskip -1.5cm +\frac{\Delta_{nj}}2  \, \Big[\, d\,\kappa_j
(V_j-V_{j+1}) \, u^*_j + \sum_{j'> j} 2^{j-j'}\big(N
_{nj'}-N_{nj}\big )\Big] \,,%%
\end{eqnarray} \end{subequations}%%%%
where we omit tildes in the new variables. Equations \Ref{rMSM}
represent the\emph{ MZS model for TBL near flat plane}.

Unlike Eqs. (\ref{eq:iMSM}), in Eqs.~\Ref{rMSM} the zone and scale
indices, $j$ and $n$, can be both  positive and  negative: thus,
$V_0$ corresponds to the velocity of the scale $\delta$, $V_{-1}$
-- to the structures of the scale $2\delta$, and so on.

We can use general formulas \Ref{expV-PM-a}  to reconstruct the
mean velocity profiles, $\bm V\SPM(\bm\rho)$. In this case the
expression for $\bm V\SPM(\bm\rho)$ take on especially simple
form, since  all functions $\bm\Phi_j(\bm\rho)$ already have a
scale-invariant form. Taking into account our rescaling to the
``inner'' units, we can write:%%
\bea\label{eq:V-tbl}
 \bm V\SPM(y) = {\rm Re}\Big[\sum_jV_j\Phi_{\rm
 un}(2^j\pi^2 y/2 \delta)\Big] \,, %%
 \EEA%%
where $y$ is the distance to the wall, $\Phi_{\rm un}(\xi)$ is
given by \Ref{univ-b} and $\bm V\SPM(y)$ is measured in the units
$U_\tau$.

%%%%%%%%%%%%%%%%%%%%%%%%%%%%%%%%%%%%
%%
\section{\label{s:sol}%%
Solution of the MZS equations}
\subsection{\label{ss:laminar}%%
Laminar velocity profile and its instability}%%
\subsubsection{Comparison of the full and \PM-laminar profiles}%%
 The simplest solution of MZS-equations~\Ref{MSM}, in
which all $u_{nj}=0$,   corresponds to the laminar flow: %%
\bea \label{eq:Vj-lam}%%
 V_j = V_{j}^0 \equiv \frac1{\Gamma_j}
= \frac{\RE}{G_j\kappa_j^2} \ . \EEA%%
Using \REF{expV-PM-a}, one reconstructs the laminar profile of the
\PM-velocity %%
\bea\label{eq:Vy-lam} %%
\bm V\SPM^0(\bm\rho) = \sum_j V_j^0\bm\Phi_j'(\bm\rho)= \RE\,
\sum_j \frac{\bm\Phi_j'(\bm\rho)}{G_j\kappa_j^2}   \ .%%
\EEA%%
The laminar  profile of the full velocity, $\bm V^0(\bm\rho)$,
satisfy the linear NSE, which in the dimensionless form~\Ref{norm}
reads%%
\bea\label{eq:NSE-1}%%
 \Delta \B V^0
(\bm\rho)  + \hat {\bf x}\, \RE  =0 \ .%%
\EEA%%
According to \REF{V-PM} the \PM-velocity is understood as the
\PM-projection of the full velocity: $\hat {\B P}\SPM \{\B V(\B
\rho)\} =  \B V\SPM (\B \rho)$. Therefore, we expect that
 the laminar \PM-profile~\Ref{Vy-lam}
satisfy the equation, similar to \Ref{NSE-1}:%%
\bea\label{eq:NSE-2} %%
\hat {\B P}\SPM \{\Delta \B V \SPM^0(\bm\rho)\}
  +\hat {\bf x}\,\RE=0\ .%%
 \EEA%%
The proof of this equation is given in Appendix~\ref{a:PM}.

\begin{figure} %%
\epsfxsize=8.6 cm \epsfbox{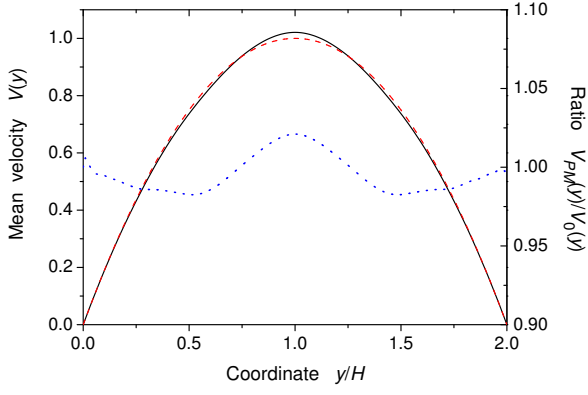} %%
\caption{\label{f:lam-profiles} Comparison of the exact (solid
line) and $PM$- (dashed line) laminar profiles for the channel
flow. The ratio of two profiles is shown by dotted line. Its
values are marked on the right axis.}
\end{figure}

In the case $\hat {\B P}\SPM \Delta= \Delta \hat {\B P}\SPM $ Eqs.
\Ref{NSE-1} and \Ref{NSE-2} coincide,  because $ \hat {\B P}\SPM
\B V \SPM=\B V \SPM$. If so,  the functions $\B V^0(\B \rho)$ and
$\B V^0\SPM(\B \rho)$ have to coincide due to uniqueness of
solutions of \REF{NSE-1} with zero boundary condition. However,
due to incompleteness of the \PM-basis, $\hat {\B P}\SPM \Delta\ne
\Delta\hat {\B P}\SPM$ and hence  the profiles $\B V^0(\B \rho)$
and  $\B V^0\SPM(\B \rho)$ are expected to differ.  In other
words, \REF{Vy-lam}  cannot reconstruct \emph{exactly} the laminar
profile $\B V^0(\B \rho)$. This is the price to pay for the
incompleteness the \PM-basis. Nevertheless the \PM-basis is ``full
enough" to allow the reconstruction of any ``physically possible"
mean velocity profile with good accuracy. As a first demonstration
of this fact we compare in Fig.~\ref{f:lam-profiles} the exact and
the \PM-\emph{laminar} profiles for channel flow, which differ
only in few percents. This small loss of accuracy  is
insignificant with respect to a dramatic simplification of the
calculation scheme for $\B V(\B \rho,t)$: for large $\RE$ the mean
velocity field in \PM-representation has $N\approx \ln \RE $
significant coefficients $V_j$, while in the corresponding
complete cell basis one has to account for $\sim\RE\gg N$
functions.

%%%%%%%%%%%%%%%%%%%%
%%%%%%%%%%%%%%%%%%%%

%%
\subsubsection{Instability of the laminar flow at $\RE=\RE\sb{\rm cr}$}%%
The laminar solution~\Ref{Vj-lam} exists for all $\RE$. For large
$\RE$, however, it becomes unstable with respect to excitation of
turbulent near-wall eddies $u_j$. A simple analysis of the
linearized \REF{MSM-b} (i.e.  $\C N_{nj} \Rightarrow 0$) shows
that the  instability condition of  $j$-eddy reads: %%
\bea\label{eq:crit}%%
 \gamma_j <  d\,(V_j-V_{j+1})/ {2\sigma_j} \ .%%
\EEA%%
Substituting here  $V_j$ from ~\REF{Vj-lam} one gets:
\[
\frac{g_j\kappa_j^2}{\RE} < \frac{d}{2\sigma_j}
\Big(\frac{1}{G_j\kappa_j^2}-\frac{1}{G_{j+1}\kappa_{j+1}^2}\Big)\RE
\,,\]%%
which can be rewritten as the condition for $\RE$:%%
\bea\label{eq:Rej} %%
&& \hskip -1.5cm \RE > \RE_{j}\equiv  \kappa_j^{3/2} \sqrt{
\frac{\sigma_j}{s_j} \frac{ g_j\,G_j\,G_{j+1}} {\, d\,(G_{j+1}
-G_j\,\kappa_j^2/\kappa_{j+1}^2 )} } \ .%%
\EEA%%
 It is clear that $\RE_j \propto \kappa_j^{3/2}$, therefore,
 usually the first unstable zone is the first one, $j=1$.

We want to stress, that \REF{Rej} gives a correct  instability
threshold only for the {\em first} unstable zone:%%
\bea\label{eq:Re1}%%
\RE\sb{cr}= \RE_{1}= \kappa_1^{3/2} \sqrt{ \frac{\sigma_1\,
g_1\,G_1\,G_{2}} {s_1\, d\,(G_{2} -G_1\,\kappa_1^2/\kappa_{2}^2 )}
} \ .%%
\EEA%%
The excitation of turbulence in, say, first zone will lead to a
significant momentum flux from the 1st to the 2nd zone. As a
result, the mean velocity $V_2$ will increase, the 2nd zone
velocity gradient $V_2-V_3$ will increase too, and  the
instability condition for $u_2$ will be therefore satisfied for
smaller Reynolds numbers  than predicted by \REF{Rej}.  We will
show later, that the real instability threshold for $j$-th zone is
proportional to $\RE_j\propto\kappa_j$, i.e. is much smaller than
the ``laminar'' result~\Ref{Rej}.

\subsection{\label{ss:abc=0}%%
Wall-bounded flows in the approximation of  near-wall eddies
($a=b=c=0$).} %%
In the previous Sec.~\ref{ss:laminar} we have found the laminar
solution of  the MZS-Eqs.~\Ref{MSM} with  $u_{nj}=0$ and have
shown that this solution becomes unstable at $\RE=\RE\sb{cr}$ with
respect to excitation of the near wall velocity $u_1\equiv
u_{11}$. As the next step in this Section we analyze a much more
general solution of Eqs.~\Ref{MSM}, that allows  non-zero values
of all near wall velocities, $u_{jj}\equiv u_j$. The $jj$-eddies
can be excited by a direct interaction with the \PM-velocities,
$V_j$. To prohibit the turbulent cascades, leading to excitation
of other $nj$-eddies (with $n>j$), one can neglect in \REF{MSM-b}
the interaction term $\C N_{nj}$ (i.e. put $a=b=c=0$). In this
case the MZS-Eqs.~\Ref{MSM} take on a simple form:
\begin{subequations}\label{eq:a0}
\begin{eqnarray}\label{eq:a0-a}
&& \hskip -1.5cm  \frac{dV_j}{dt} = -\Gamma_jV_j +1
   +d\,\kappa_j\left(u_{j-1}^2-u_j^2\right)
\,,\\\label{eq:a0-b}
 && \hskip -1.47cm  \frac{du_j}{dt} = -\gamma_ju_j
    +\frac{d}{2\sigma_j}\left(V_j-V_{j+1}\right)u_j^*
\,.\end{eqnarray}
\end{subequations}
In the stationary regime ($d/dt=0$) only the finite number
$(m-1)\ge 0$  of turbulent velocities are non-zero, i.e., $ u_j =
0$ for $j\ge m$. This number  depends on $\RE$. As it follows from
Eq.~(\ref{eq:a0-a}), the \PM-velocities $V_j$ for $j>m$ coincide
with the laminar ones:%%
\bea\label{eq:V-large-j}
  V_j = V_j^0=\Gamma_j^{-1} \sim  2^{-2j}\RE\,,
  \quad j>m=m(\RE)\ . %%
  \EEA%%
In the unstable region $j<m$, as it is follows from~\REF{a0-b},
the zone velocity difference, $\Delta V_j\equiv V_j-V_{j+1}$
coincide with it critical value, see \REF{crit}:%%
\bea\label{eq:DVj} %%
\Delta V_j&=&\Delta V_j^{\rm cr }=\frac{2 \gamma_j\sigma_j}{
d}=A_j/\RE\,,\br
  A_j& \equiv & 2 g_j\kappa_j^2 \sigma_j\big/  d\ .%%
 \EEA
This allows one to find all $V_j$ via the last velocity in the
unstable region, $V_m$:%%
\bea\label{eq:aVj}%%
V_j&=&V_m + (B_m-B_j)/\RE\,,\br%%
B_j &\equiv&  \sum_{i=1}^{j-1} A_i \ . %%
\EEA%%
Multiplying \REF{a0-a} by $s_j$ and summing up from $j=1$ to
$j=m$, one finds the equation for $V_m$ with the solution:%%
\bse\label{eq:sVm}\bea\label{eq:sVm-a} %%
V_m &=& \frac{1-\sigma_{m+1}}{C_m}\RE -\frac{D_m}{C_m\RE}\,, \\
\label{eq:sVm-b} %%
C_m &\equiv& \sum_{j=1}^m s_j G_j\kappa_j^2
    \,\sim \, 2^m
\,,\\ \label{eq:sVm-c}
  D_m &\equiv& \sum_{j=1}^m s_j G_j\kappa_j^2(B_m-B_j)
    \,\sim \, 2^{2m} \ . %%
\EEA \ESE%%
Note, that in spite of the large number of different parameters
($A_m,\dots,D _m$), all of them are provided with explicit
expressions and can be easily evaluated.

Now one finds the total linear mechanical momentum $\C P$ (i.e.
the total flux of the fluid) of the flow from its
definition~\Ref{mot-ints-b} and Eqs.~\Ref{V-large-j} and
\Ref{sVm-a} for $V_j$  for $j\le m$ and for $j>m$:%%
\bse\label{eq:pi}
\begin{eqnarray}\label{eq:pi-a}
  \mathcal P &=& \sum_{j=1}^ms_jV_j +\sum_{j=m+1}^\infty s_jV_j^0
\br  &=&    E_m\RE + F_m/ \RE\,,  \\ \label{eq:pi-b}%%
E _m &\equiv& \frac{(1-\sigma_{m+1})^2}{C_m}+\sum_{j=m+1}^\infty
\frac{s_j}{G_j\kappa_j^2}\,,\br%%
 F_m &\equiv&
\sum_{j=1}^ms_j(B_m-B_j) -\frac{(1-\sigma_{m+1})D_m}{C_m}\ .
\end{eqnarray} \ESE%%

So far we treated the last index $m$ in the unstable zones as
fixed. However, it depends on the Reynolds number $\RE$ and is
defined from the condition
\[
  V_m -V_{m+1} <\Delta V_m^{\rm cr} =\frac{A_m}\RE\,,
\]%%
which ensures that the turbulent velocity $u_m$ will not be
excited, $u_m=0$. This condition can be written as%%
\bea\label{eq:Rem}
  \RE^2 < \frac{A_mC_m+D_m}
   {1-\sigma_{m+1}-C_m/(G_{m+1}\kappa_{m+1}^2)}
    \equiv \RE_m^2 \ .%%
    \EEA%%
Actually, it will be exactly $(m-1)$ excited turbulent velocities
$u_j$, if
\begin{eqnarray}\label{eq:diapason}
  \RE_{m-1} < \RE < \RE_{m}
\,.\end{eqnarray} %%
The analysis shows that this condition selects the index $m$, for
which the total momentum $\mathcal P$ (\ref{eq:pi}) has minimum
value.

\begin{figure}%%
\epsfxsize=8.6cm \epsfbox{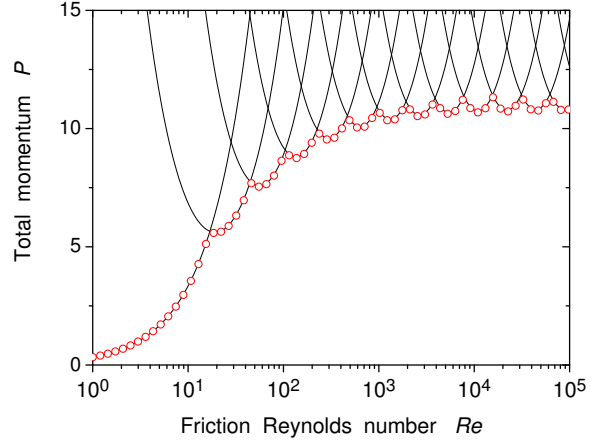}%%
\caption{\label{f:no-turbulent} Total momentum of the flow {\it
vs.} the Reynolds number $\RE$. Circles denote   numerical data
for the channel flow, $d=0.046$; solid lines - the analytical
prediction (\ref{eq:pi}) for different critical zone numbers $m$,
$m=1,2,\dots$ (from left to right).}
\end{figure}

Let us analyze the dependence~\Ref{pi-a} of the total momentum in
the limit of extremely large Reynolds numbers, $\ln \RE\gg 1$,
and, respectively, large critical zone numbers $m$. In this case
 we can use  the scale-invariant limit for all parameters:%%
\bea\label{eq:lim} E_m = 2^{-m}\tilde E\,,\quad
 F_m = 2^{m}\~F\,, \quad \RE_m = 2^m\tilde R\,,%%
 \EEA%%
 where $\tilde E$, $\~ F$ and $\~ R$ are some geometry-dependent
constants. Now for Reynolds numbers inside the range
  (\ref{eq:diapason}) we can write in \REF{pi-a} $ x(\RE ) = \RE/
(2^m\tilde R)$, where $1/2<x(\RE)<1$. Then $\mathcal P$   for
large enough $\RE$ [in the region of validity of the scale
invariant limit~\Ref{lim}],  is given by%%
\bea \label{eq:P-x}%%
  \mathcal P (\RE)  = x(\RE)\,\tilde E\,\tilde R
  +\frac{\tilde F}{x(\RE)\,\tilde R }\ . \EEA%%
Since $x(\RE)$ is bounded between 1/2 and 1, the flux~\Ref{P-x} is
bounded from above by one of the constants%%
\[ %%
 \mathcal P (\RE) \le  \max \Big \{ \frac {\tilde E\,\tilde
 R}2+
 \frac{2 \tilde F}{\tilde R }\,, \
 {\tilde E\,\tilde R} +\frac{\tilde F}{\tilde R }\Big \}\
.
\]%%
and {\em does not grow infinitely for infinite Reynolds numbers}.
In other words, neglecting the dissipation of energy in turbulent
cascades, one concludes that with the given cross-section area of
the flow and  pressure gradient, the total flux of the fluid
reaches some limit in spite of  the infinite decrease of the
kinematic viscosity. The numerical and analytical calculations for
the channel geometry, shown in Fig.~\ref{f:no-turbulent}, support
this unexpected conclusion.

%%%%%%%%%%%%%%%%%%%%
%%%%%%%%%%%%%%%%%%%%

\begin{figure}%%
\epsfxsize=8.7cm \epsfbox{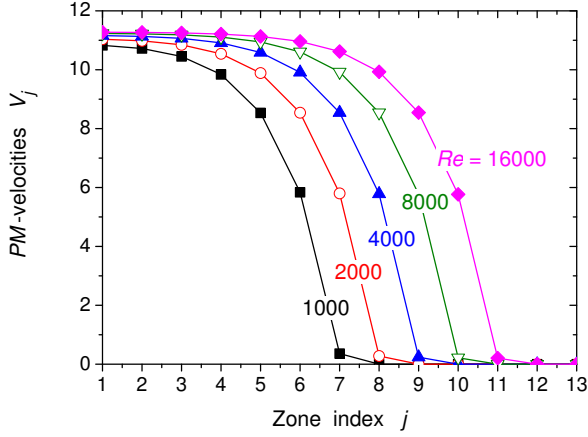}%%
\caption{\label{f:no-turbulent-Vj} Mean zone velocities $V_j$ {\it
vs.} the zone index $j$ for different Reynolds numbers $\RE$.}
\end{figure}

The reason for this strange  behavior is that the stationary zone
velocity differences,  $\Delta V_j$, in the unstable region $j<m$
are determined solely by the dissipation of turbulent eddies
$\gamma_j$, \REF{DVj},  that go to zero as $\RE\to\infty$ (for
finite $j$). Thus, in the limit $\RE\to\infty$,  $\Delta V_j\to0$
and $V_j\to$ const.  This conclusion is illustrated  in
Fig.~\ref{f:no-turbulent-Vj}. Clearly,  for large $\RE$ first few
velocities remain unchanged, and the only effect of increasing
$\RE$ is the shift of the ``dissipative cutoff'' towards the
smaller scales. However, the total momentum is determined mainly
by the first few zones and thus remains the same.

%%%%%%%%%%%%%%%%%%%%
%%%%%%%%%%%%%%%%%%%%

This analysis shows importance of the turbulent cascade for the
experimentally observed  characteristics of real flows, i.e.
infinite increase of the total flux, logarithmic profiles, etc.
The stationary condition $\Delta V_j=\Delta V_j\sp{cr}$ will hold
even if one accounts for a turbulent cascade, which adds some {\em
turbulent damping} $\gamma\Sp T_{j}$ for turbulent near-wall
eddies instead of usual one $\gamma_j$. The turbulent damping {\em
does not vanish in the limit $\RE\to\infty$}. Thus, one obtains
$\Delta V_j=\Delta V_j^{\rm cr} \ne0$, and the total momentum will
infinitely increase, as it is shown in the following
Sec.~\ref{ss:log}.

\subsection{\label{ss:log}%%
Wall bounded flows in the ``turbulent-viscosity" approximation} %%
 In this Section we show that even rough account for the
turbulent cascade in the MZS-model already gives qualitatively
correct analytical results.

\subsubsection{\label{sss:tv-appr}%%
MZS equations in the turbulent viscosity approximation}

In the fully developed turbulent regime, the  action of the
interaction term $\mathcal N_{nj}$ on the eddy $u_{nj}$ can be
approximately accounted for in the ``turbulent viscosity"
approximation in which the energy flux from the energy containing
$jj$-eddies toward small scales in the $j$-zone is replaced by a
nonlinear damping term, ensuring the same loss of their energy.
Formally this can be done by replacing the nonlinear term $\C
N_{jj}$ in the full MZS-Eqs.~\Ref{rMSM} by some effective
turbulent damping term, $\gamma_{j}\Sp T$:%%
\bse  \label{eq:t-dam} \bea \label{eq:t-dam-a}%%
\mathcal N_{jj}  &\Rightarrow& -\gamma_{j}\Sp T u_{j} \,,\qquad
\gamma_{j}\Sp T =  \alpha\,  \kappa_j|u_{j}|\,,%%
 \\ \label{eq:t-dam-b}\alpha &\approx&  (a-c)\ .%%
\EEA\ESE %%
Here $\gamma_{j}\Sp T$, is chosen as the turnover frequency of
$jj$-eddies, $\kappa_j|u_{j}|$ with some dimensionless prefactor
$\alpha$. This prefactor is evaluated in \REF{t-dam-b} by equating
the total rate of energy dissipation in the model system with the
effective turbulent damping~\Ref{t-dam-a} and the energy flux
toward small scales in the full shell model, see, e.q.
Ref~\cite{98LPPPV}.

In the suggested  {\em effective turbulent damping} approximation,
the MZS-Eqs.~\Ref{rMSM}  take the form
\begin{subequations}\label{eq:turb-damp}
\begin{eqnarray}\label{eq:turb-damp-a}
\frac{dV_j}{dt} &=& -\Gamma_jV_j +1
+d\,\kappa_j\left(u_{j-1}^2-u_j^2\right) \,,\br%%
\frac{du_j}{dt} &=& -(\gamma_j+\alpha \kappa_j|u_j|)u_j +\frac{
d}{2\sigma_j} \left(V_j-V_{j+1}\right)u_j^* \ . \\
\label{eq:turb-damp-b}%%
 \end{eqnarray}
\end{subequations}
These equations  are different from the MZS-equations in the
near-wall-eddy approximation, \REF{a0}, by the only term $\propto
\alpha$.

For large $\RE$  in the near wall region  \REF{turb-damp}
 have a more simple,  scale-invariant form, in which all
geometry dependent factors are taken in their small scale
limit~\Ref{as-con}:
\begin{subequations}\label{eq:turb-d}
\begin{eqnarray}\label{eq:turb-d-a}
\frac{dV_j}{dt} &=& -\frac{G\kappa_j^2 }{\RE}  V_j +1 +d\,\kappa_j
\left(u_{j-1}^2- u_j^2\right) \,,\br%%
\frac{du_j}{dt} &=& -\Big (\frac{g\kappa_j }{\RE} +\alpha
|u_j|\Big )\kappa_j u_j +\frac{d\,\kappa_j
}{2} \left(V_j-V_{j+1}\right)u_j^* \,,  \\
\label{eq:turb-d-b}%%
 \end{eqnarray}
\end{subequations}
where $\kappa_j = 2^j/2 s\ .$
\subsubsection{\label{sss:in-int}Inertial interval solution}
In this section we consider the stable stationary solution of
\REF{turb-damp}. This solution is real.

For  $\log_2 \RE\gg 1$, in the inertial interval of scales, all
needed information contained in  \REF{turb-damp-a} may be obtained
from \REF{p-in} for the momentum flux, $\F p_j$, in which one can
neglect the damping term $\propto 1/\RE$. This gives
\bea\label{eq:uj}%%
 \F p_j = \frac{d u_j^2}{2}  = \sum_{j'=1}^js_{j'}=1-\sigma_{j+1}\,,%%
\EEA%%
where $\sigma_j$ is defined by \REF{nots-c}. The substitution of
$u_j$ from \REF{uj} in \REF{turb-d-b} gives  an expression for
$V_{j+1}$ via $V_j$. This allows one to find $V_j$ for any $j$
outside the viscous region via some constant $V_0$:%%
\bea\label{eq:set}%%
V_j=V_0-\frac{\sqrt2\alpha}{d^{3/2}}
\sum_{i=1}^j\frac{\sigma_i\sqrt{1-\sigma_{i+1}}}{s_i} \ .%%
\EEA%%
 For  $j>j_*$, (where $j_*$ is equal,
say, 3) one can take in \REF{set}  $\sigma_j=2\, s_j\ll  1$. One
concludes, that in the inertial interval the \PM-velocities
decrease linearly in $j$:%%
\bse \label{eq:set1} \bea\label{eq:set1a}%%
V_j= V_0+\Delta -\frac{2\sqrt2\,\alpha j }{d^{3/2}}\,,\quad
j>j_*\,,
\\ \label{eq:set1b}%%
\Delta=  \frac{\sqrt2\alpha}{d^{3/2}}\sum_{i=1}^\infty
\left[2-\frac{\sigma_i\sqrt{1-\sigma_{i+1}}}{s_i} \right]\ . %%
\EEA \ESE%%
Here the geometry dependent constant $\Delta $ was found by
comparison of Eqs. \Ref{set} and \Ref{set1}, and for different
geometries it evaluates to:%%
\bea \Delta \approx \frac{\alpha}{d^{3/2}}\times \Big \{
\begin{array}{c@{\qquad}c}
2.7\,, & \text{channel}\,,\\
2.4 \,, & \text{pipe}\ .
\end{array} \EEA%%

The equation \Ref{set1} can be directly obtained from the
condition of the constancy of the momentum flux~\Ref{const-p},
that gives:%%
\bea\label{eq:tmp-3-1}%%
 u_j =\sqrt{2/d} =\text{const.}%%
\EEA%%
Then, in the stationary case, \REF{turb-d-b} determines the
$j$-independent difference $V_j-V_{j+1}$, necessary for keeping
the amplitudes of the near-wall eddies at the constant level. This
agrees with \REF{set1}.

In the derivation of Eqs.~\Ref{set} and \Ref{set1} we cancelled in
\REF{turb-damp-b} $u_j=u_j^*$, assuming that $u_j\ne 0$. At a
given $\RE$ this assumption is valid only for $j\le j_0$.  Here
$j_0$ is the zone index of the last unstable zone, for which in
\REF{turb-d-b} $d(\ln{u_j})/dt>0$ at $u_j\to 0$. To find $j_0$ we
consider  \REF{turb-d-b} as an equation for a continuous  index
$q$ and set its RHS to zero with $u_q=0$: %%
\bse \label{eq:cond}  \bea \label{eq:cond-a} %%
\frac{g\kappa_q}{\RE}&=&\frac{d}{2}(V_q-V_{q+1})=
\frac{\sqrt2\alpha}{\sqrt{d}}
\quad  \Rightarrow\\ \label{eq:cond-b}  %%
q(\RE )&=&\log_2\RE+\log_2\left[2\sqrt2\alpha s
\big/g\sqrt{d}\right]\ .%%
\EEA\ESE%%
The index $j_0$ is then the integer part of $q$.

The solution~\REF{set} has still an unknown constant, $V_0$. This
constant can be found from the stationarity
condition~\Ref{const-p},%%
\bea\label{eq:cond0} \F p^-=\sum _{j=1}^{\infty} s_j \Gamma_j
V_j'=\F p^+=1 \,, %%
\EEA%%
 requiring that the total influx of the linear momentum,
caused  by the pressure gradient, must dissipate at the wall, due
to the viscous friction. The product $s_j \Gamma_j\sim 2^{j}/\RE$.
Therefore, the sum in \REF{cond0} is dominated by last few terms
with $j\approx j_0\gg j_*$. This allows one to use a more simple
\REF{set1} instead of \REF{set}:%%
\bse \label{eq:set3} \bea\label{eq:set3a}%%
1&=&  \frac{G}{4\RE\,s} \sum_{j=1}^{j_0}2^j\left[V_0+\Delta
-\frac{2\sqrt2\,\alpha\, j }{d^{3/2}}\right] \\
\label{eq:set3b} %%
&\approx&  \frac{G}{2\RE\, s} 2^{j_0} \left[V_0+\Delta
-\frac{2\sqrt2\,\alpha j_0}{d^{3/2}}\right] \\
\label{eq:set3c}  %%
&\approx& \frac{\sqrt2\, \alpha}{\sqrt{d} }\frac{G}{g}
 \left\{V_0+\Delta
-\frac{2\sqrt2\,\alpha \, q (\RE)}{d^{3/2}}\right\}\ .%%
\EEA\ESE%%
Together with \REF{set1a} this gives the  dependence of $V_j$ on
$j$ for the near wall region, that is \emph{geometry independent}:
\bea\label{eq:V1}%%
 V_j= \frac{2\sqrt2 \alpha}{
d^{3/2} } \big[q(\RE)-j\big]+ \frac{\sqrt d} {\sqrt2\alpha}\frac g
G\,, \quad j>3\ .\EEA %%
The only $\RE$-dependent factor here is the position of the
viscous cutoff $q(\RE)$, given by the \REF{cond-b}.  Counting $j$
from $q(\RE)$, one has $\RE$ independent \PM-velocities.  In the
physical space this corresponds to the \emph{universality of the
TBL profile} measured in the ``wall units" $y^+=y\, \RE$.

\begin{figure*}%%
\epsfxsize=8.6cm \epsfbox{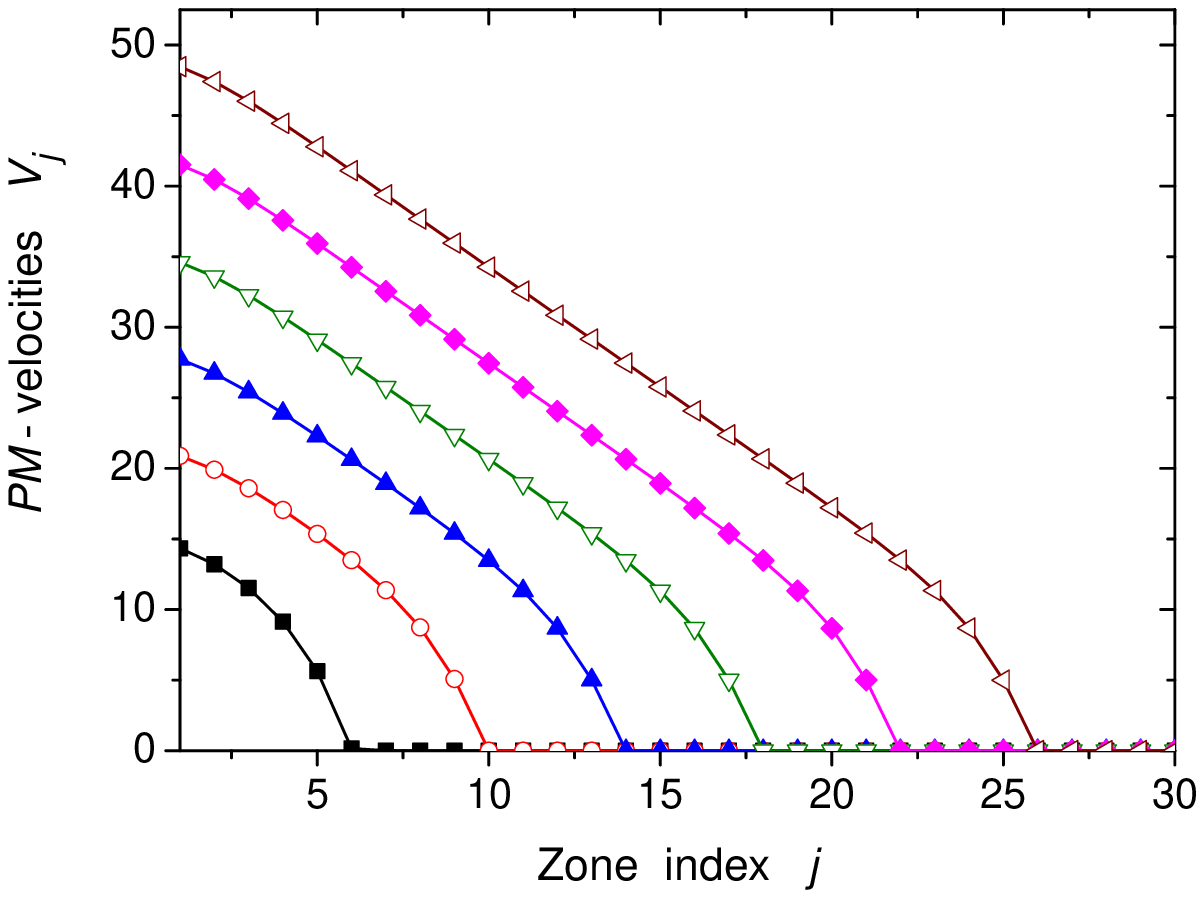} %%
\hskip -1cm \epsfxsize=9.4 cm\epsfbox{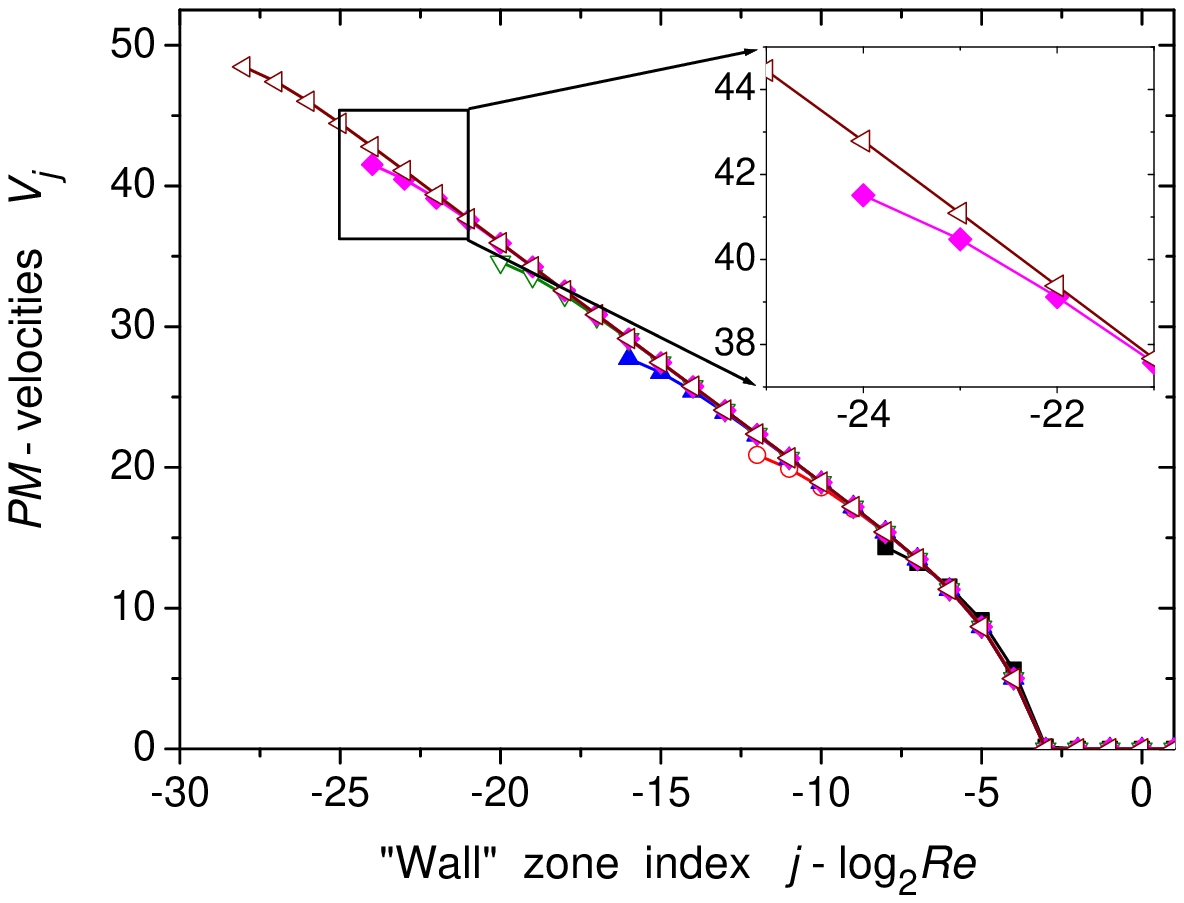}
 \caption{\label{f:graph4}%%
\PM-velocities $V_j$  for different Reynolds numbers $\RE$ as a
function of the zone index $j$  (left panel) and of the
``near-wall'' zone index $(j-\log_2\RE)$ (right panel). The lines
from lower to upper correspond to  $\log_2\RE=9,13,17,21,25$  and
 29, respectively.}
\end{figure*}

Notice, that in the derivation of \REF{set}, we neglected in
\REF{turb-d-b} the viscous damping  term, $g\kappa_j^2/\RE$, with
respect to the turbulent one, $\alpha|u_j|\kappa_j$. This
approximation fails near the viscous cutoff, because
$g\kappa_j/\RE \propto 2^{j}$ increases exponentially toward large
$j$, while $\alpha|u_j|$ is approximately constant. More detailed
analysis shows, that \REF{set} fails only for last two $j$ before
the cutoff and instead of sharp cutoff at $j=j_0$ there is a soft
decrease of $V_j$ in  two zones near $j=j_0$, see
Fig.~\ref{f:graph4}.

One can easily see, that the linear in $j$ set of zone
velocities~\Ref{V1} in the inertial interval corresponds to the
log-profile of the mean velocity in the physical space:
$V(y)\sim\ln (y\RE /L)$. Actually, by a direct calculation one
shows, that the log-profile%%
\bea\label{eq:K} %%
V\sp{log}(y) = \frac{1}{\kappa\Sb K}\ln\Big(\frac{y\,\RE}{L}\Big)
+B \,,%%
\EEA%%
corresponds to  the zone velocities
\begin{eqnarray}\label{eq:log}
V\sp{log}_j&=& \frac{\ln2}{\kappa\Sb K}\left(\log_2\RE-j\right)
+\left(B-\frac{1+\gamma\Sb E-\ln (4/\pi)}{\kappa\Sb K}\right)
\nonumber\\ &\simeq& \frac{0.69}{\kappa\Sb K}
\left(\log_2\RE-j\right) +B-\frac{1.34}{\kappa \Sb K}
\,,\end{eqnarray}%%
 where $\gamma \Sb E\simeq 0.58$ is the Euler gamma
constant.

Thus, the MZS model describes the transition to the universal
log-profile with the von-Karman constant%%
\bea\label{eq:Kar}%%
 \kappa\Sb K  = \frac{\ln2\,}{2\sqrt2}\frac{d^{3/2}}{\alpha}
 \simeq 0.25\frac{d^{3/2}}\alpha \ .%%
 \EEA
 %%
%%%%%%%%%%%%%%%%%%%%%%%%%%%%%%%%%%%%%%
\subsection{\label{ss:numerics}%%
Numerical analysis of the turbulent channel flow in the
 MSZ-model}%%
%%
\begin{figure}%%
\epsfxsize=8.6cm \epsfbox{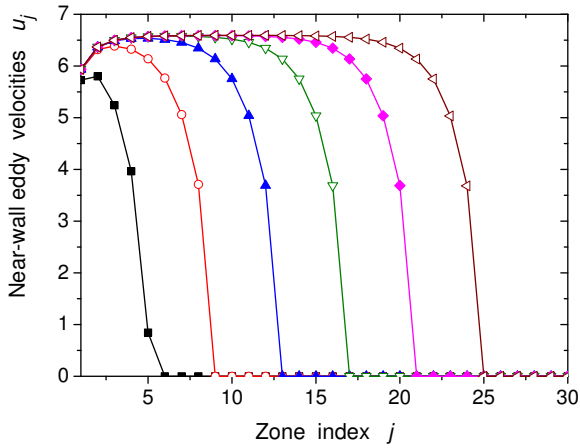}%%
\caption{\label{f:graph6}%%
Magnitude of the near-wall turbulent eddies $|u_j|$ as a function
of the zone index $j$ for different $\RE$. The lines are marked as
in Fig.~\ref{f:graph4}. }
\end{figure}%%
 This Section is devoted to the numerical analysis of the MZS-model of
the channel flow in the effective turbulent damping approximation,
\REF{turb-damp}. For simplicity we adopt $g_j=G_j$.
\PM-damping factor $G_j$ was found according to \REF{res-2} as the
matrix elements of the viscous operator on the real part of the
\PM-basis $\Phi_j'(y)$. This gives%%
\[%%
G_j =  2^{j+2} s_j\ . %%
\]%%
Parameters $d=4.6\times 10^{-2}$ and  $\alpha = 5.9\times10^{-3}$ were
chosen to reproduce the  experimental values of the  universal constants
$\kappa\Sb K\approx0.4$ and $B\approx 5.2$.%%

In numerical analysis,   $30$ zones were sufficient to describe
the flows with  Reynolds numbers up to $10^9$. In order to find
the stationary solution of the full MSZ-Eqs.~\Ref{turb-damp} we
develop extremely stable and efficient iteration procedure, based
on the essential physics of the problem, see
Appendix~\ref{a:iter}. In spite of the huge Reynolds numbers, the
accuracy better than $10^{-6}$ was reached with about 100-200
iteration. Actually, using our approach (MZS-model and the
iteration procedure) one can simulate turbulent wall-bounded flows
for {\em arbitrary large} $\RE$ with very modest PC, even with
XT486@40MHz, 8Mb RAM.%%
\subsubsection{\label{eq:zone}%%
Behavior of the zone velocities $V_j$ and $u_j$}%%

In Fig.~\ref{f:graph4}, left panel,   we plot \PM-velocities $V_j$
for different $\RE$ from 500 to 5$\times 10^8$. One clearly sees
the inertial interval, where $V_j$ decrease linearly with $j$ in
agreement with \REF{set}. As we mentioned above there is a ``soft"
viscous cutoff, that involves last 2 zones.
\begin{figure}%%
\epsfxsize=8.6cm \epsfbox{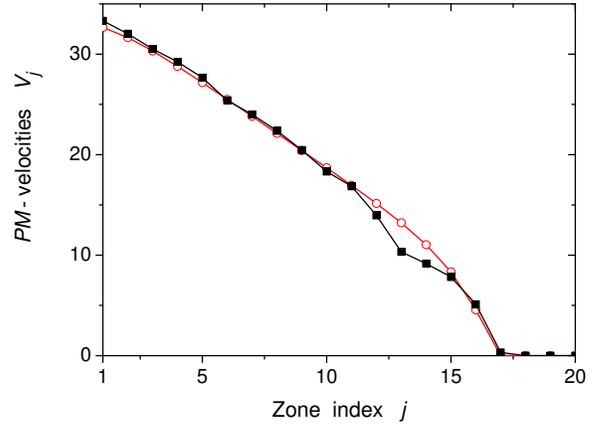}%%
\caption{\label{f:Vj-Sabra} Comparison of numerical solutions of
the full MZS Eqs.~\Ref{rMSM} [black squares] and MZS
Eqs.~\Ref{turb-damp} in the approximation of turbulent
viscosity~\Ref{t-dam} [empty circles]. $\RE=10^6$.} %%
\end{figure}%%
In order to demonstrate the phenomenon of universality, in the
right panel of  Fig.~\ref{f:graph4}  we re-plot the same
velocities $V_j$ as function of the ``near-wall'' zone index
$j-\log_2\RE$ as it is suggested by \REF{V1}. There is a perfect
collapse of all lines. Importantly, they collapse not only in the
inertial interval, but also in the dissipative cutoff range. There
is a non-negligible difference for the first two-three zones,
$j=1,2$, (see the blow up  in the lower panel) that is caused by
the non-negligible momentum influx in these zones. This deviation
is described by \REF{set}.

Fig.~\ref{f:graph6} shows the  magnitudes of the near-wall
turbulent eddies, $|u_j|$, for the same set  of $\RE$'s. Again,
the numerical results are in full agreement with our theoretical
understanding: the magnitudes of the near-wall eddies in the
inertial interval are constant, as it required by \REF{tmp-3-1}
from the condition of the constancy of momentum flux.   At the
dissipative cutoff we observe a  decrease of $u_j$ involving 3--4
zones.  A small decrease of $u_j$ at the first few shells is
caused by the non-zero momentum flux into these zones; in this
region $u_j$ with good accuracy can be found from%%
\bea\label{eq:start}%%
 \mathfrak p_j =  du_j^2/2  = 1-\sigma_{j+1}\ .
\EEA To check how the approximation of turbulent viscosity,
\REF{t-dam},  affects  the resulting mean velocity profile we
compare in Fig.~\ref{f:Vj-Sabra} the solutions of the full MZS
Eqs.~\Ref{rMSM} (black squares) and MZS Eqs.~\Ref{turb-damp} in
the approximation of turbulent viscosity, \REF{t-dam} (empty
circles). These solutions practically coincide except for three
zones ($j=12,13,14$) near the viscous cutoff. In these zones just
a few shells are excited in the full model~\Ref{rMSM} and the
observed difference is an artefact of discreetness of the scale
space in the shell models (recall, that in the shell
model~\Ref{rMSM} the spacing parameter $\lambda=2$). In the case
when the details of turbulent cascades near the wall are
physically important (e.g., in turbulent flows with polymeric
additives) one has to use more detailed shell representation of
turbulent velocity field, say with $\lambda=\sqrt2$ or even
$\lambda=2^{1/4}$, preserving the interaction range in the scale
space unchanged.  This modification of our model is under
construction and will be published elsewhere.

We expect that the more detailed models with $\lambda<2$ will give
the mean profile, closer to that, given by reduced  model with
turbulent damping~\Ref{turb-damp}, in which the energy dissipation
is affected by the scale spacing. In this paper we are not
interested in details of turbulent cascades and will use further
only Eqs.~\Ref{turb-damp}.%%
\subsubsection{\label{ss:reconstr-V}%%
Reconstruction of the mean velocity profile}

\begin{figure}%%
\epsfxsize=9cm \epsfbox{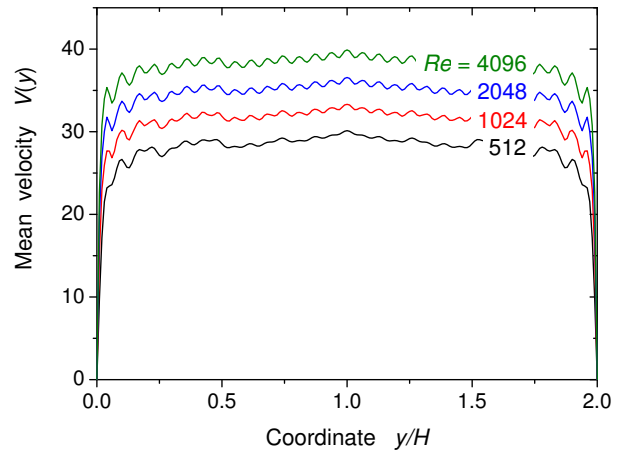}%%
\caption{\label{f:graph9}%%
Reconstructed \PM-velocity profiles $V_{PM}(y)$ for different
Reynolds numbers. }
\end{figure}

The equation~\Ref{expV-PM-a} reconstructs \PM-velocity profile
$V_{PM}(y)$ from the set of $V_j$. For the case of the channel
flow one has:%%
\bea\label{eq:recon}%%
V_{PM}(y)&=&\sum_{j=1}^\infty V_j \Phi_j'(y)\,,\br %%
\Phi_j'(y)&=&\sum_{m=2^{j-1}}^{2^{j}-1}p_{m}\phi_{2m-1}(y)\,, %%
\EEA%%
where $p_m$ and $\phi_m(y)$ for the channel geometry are given by
\REF{F-bas}.  The results of such reconstruction for a set of
$\RE$'s are shown in Fig.~\ref{f:graph9}.   We see that
reconstructed \PM-profiles show all qualitative features of the
``real'' mean velocity profiles in the well-developed turbulent
regime.  Namely, the mean \PM-velocity is almost constant in the
main part of the flow [with centerline velocity, increasing with
$\RE$ as $\ln(\RE)$]. As expected, all fall  of $V(y)$  occurs in
a small region near the walls. It is also clear, that the width of
this near-wall region decreases as $\RE$ increases, in the same
way as in real flows.

%%

%%%%%%%%%%%%%%%%%%%%
Figure~\ref{f:graph9} shows, that the \PM-profiles   have some
non-physical ``wiggles''. They originate from the incompleteness
of the \PM-basis. Indeed, the \PM-basis is constructed by
\REF{Phi-n} from the complete $\B \phi$-basis with some prescribed
$m$-dependence  for each $j$ function. As a result, the
\PM-expansion~\Ref{recon} can be understood as the Fourier
expansion in $\phi_m(y)\propto \sin(k_{2m-1}y)$ with discontinuities
of the Fourier amplitudes at the ``zone boundaries" $m=2^j$, which
produces the wiggles in the $y$-representation. This artifact of the
model can be removed by different ways. The simplest one is to add
some function $\~V(y)$, orthogonal to all \PM-basic functions (and
thus having zero momentum), with the amplitude and the shape of which
are determined from the problem of minimization of discontinuities in
the spectrum. It can be shown that the result of such ``smoothing'' is
fully acceptable for the most of purposes.

\begin{figure}%%
\epsfxsize=8.6cm \epsfbox{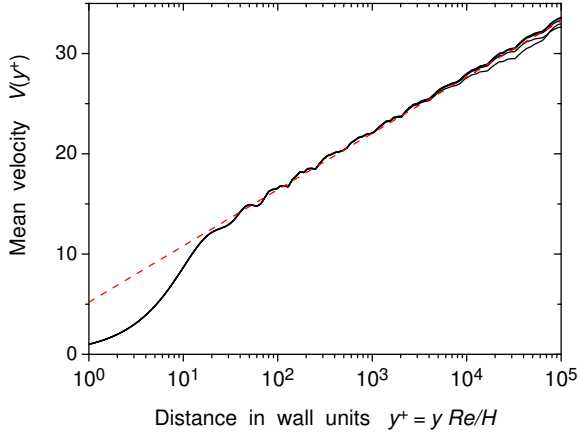}%%
\caption{\label{f:graph8}%%
Collapse of the mean-velocity profiles in the near-wall units.
Different line (from below to above) correspond to $\RE=
2^{17}\sim 1.3\times10^5, 2^{21}, 2^{25}$ and  $2^{29}\sim 5\times
10^8$ }
\end{figure}

%%%%%%%%%%%%%%%%%%%%
Figure~\ref{f:graph8} displays  the \PM-velocities  in the
near-wall region $y\ll H$ in log-linear scale for different
$\RE$'s from $\sim 1.3\times 10 ^5$ to $\sim 5\times 10^8$. The
distance from the wall is measured in near-wall viscous lengths,
$y^+=y/\delta =y\RE/H$. The collapse of profiles for different
Reynolds numbers is evident. One can see the viscous sublayer (for
$y^+\le10$) and universal log-profile (for $y^+\ge50$).

%%
\subsubsection{\label{ss:V(y)-experiment}%%
Comparison of the MZS universal mean velocity profile with
experiment and DNS}%%%%
\begin{figure}%%
\epsfxsize=9cm \epsfbox{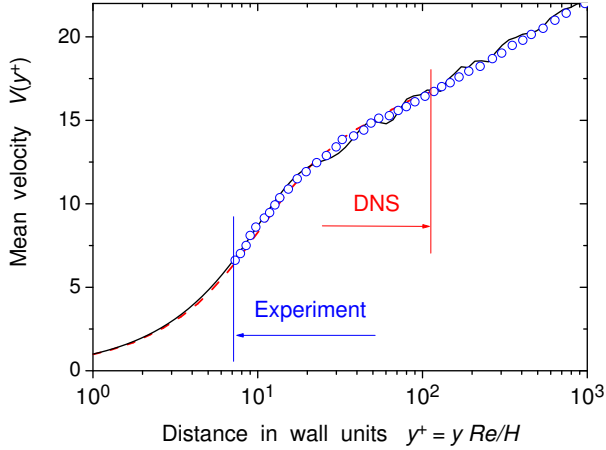}%%
\caption{\label{f:Vy-DNS-exp}%%
 Comparison of the reconstructed MZS universal mean velocity
 profile $V(y^+)/U_\tau$,
 (solid line in the  region $1\le y^+\le 10^3$ ) with the DNS results
 in channel, available in Ref.~\cite{03Rome} for $y^+<10^2$ (dashed line)
 and with the measurements in pipe, taken from Ref.~\cite{97ZS} for
$y^+>10$, (empty circles).}
\end{figure}
In Fig.~\ref{f:Vy-DNS-exp} we compared the  reconstructed MZS
universal mean velocity  with the DNS results  in
channel,~Ref.~\cite{03Rome} and with the laboratory measurements
at \Re~ up to $3.5\times 10 ^7$,  presented  in Ref.~\cite{97ZS}.
The DNS data are available for $y^+<100$, while the experimental
data are obtained for  $y^+>10$. As expected, in the overlap
region, $10<y^+<100$ both results collapse. As we explained, our
MZS model reproduce the asymptotical log-profile~\Ref{K}, and the
parameters of the model, $\alpha$ and $d$ were chosen to give
known values of $\kappa\Sb K$ and $B$, that parameterize \REF{K}.
Therefore, as expected, the MZS profile, displayed in
Fig.~\ref{f:Vy-DNS-exp} as solid line, coincide with the
experimental date in the region of large enough $y^+$. The point
is that the MZS dependence $V\SPM(y^+)$ practically coincides with
the DNS and experimental data in all region of $y^+$.  It means
that the MZS model correctly describes the basic physics that
affect the mean velocity profile in the universal near wall region
of turbulent boundary layer near the flat plane. The MZS
description of the viscous and the buffer layers \emph{does not
require adjustable parameters}.

\subsubsection{\label{ss:reconstr-E}%%
Reconstruction of the profile of the energy dissipation}%%%%
The  total dissipation rate  $\varepsilon^-_j$ in $j$-zone,
\REF{p-inf-d},  can be splited  into the  dissipation rate in the
\PM-velocity subsystem $\tilde\varepsilon^-_j$ and the dissipation
rate in the turbulent subsystem $\hat\varepsilon^-_j$:
\begin{eqnarray}\label{eq:div}
\varepsilon^-_j &=& \tilde\varepsilon^-_j +\hat\varepsilon^-_j
\,,\\\nonumber \tilde\varepsilon^-_j &=& \Gamma_j|V_j|^2\,,\qquad
\hat\varepsilon^-_j = \sum_{i\le j}\gamma_{{\rm ef},i}|u_i|^2 \ .
\end{eqnarray} %%
In the turbulent viscosity approximation the effective damping
 is given by%%
\bea\label{eq:rate}%%
 \gamma_{{\rm ef},j}=\gamma_j +\alpha\kappa_j|u_j| \ .%%
 \EEA

\begin{figure}%%
\epsfxsize=8.6cm \epsfbox{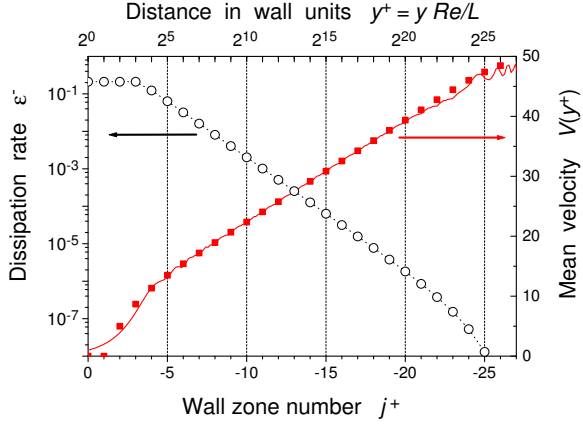}%%
\caption{\label{f:Vy-eps-y-profs} Mean velocities $V_{j^+}$ (black
squares) and turbulent dissipation rates $\ve^+_{j^+}$ (empty
circles) in the wall-zone representation $j^+$, introduced by
\REF{J}. Solid line -- reconstruction of the \PM-velocity profile
(in the wall units) $V\SPM(y^+)$ vs. distance $y^+$ in the wall
units, shown above. $\RE=5\times 10^8$. }
\end{figure}

Figure~\ref{f:Vy-eps-y-profs} displays as empty circles the
dissipation density in turbulent subsystem, normalized by the
near-wall lengthscale, i.e. $\hat\varepsilon^-_j/\RE$ as a
function of the ``wall zone number" defined here as%%
\bea\label{eq:J}%%
j^+ \equiv j-\log_2\RE+2 \ .%%
\EEA%%
The black squares in this figure show the \PM-velocities in the
``wall-zone" representation, i.e. values of $V_{j^+}$ for the same
$\RE=5 \times 10^8$. The solid line  is the log-plot of the
reconstructed (from this set of $V_j$) \PM-velocity in the
physical space, i.e. $V\SPM(y^+)$ vs. $\log_2 y^+$, shown from
above. As one sees, the solid line goes very close to the black
squares, as it should be according to the interpretation of the
\PM-velocities $V_{j^+}$ as a ``physical" velocity at some point
within $j^+$-zone, that explained in Sec.~\ref{sss:inter}. As it
is clear from Fig.~\ref{f:Vy-eps-y-profs}, the wall zone
numbers~\Ref{J} are chosen to give a very simple ``correspondence
rule": %%
\bea\label{eq:cor} %%
\text{Wall zone index}\ j^+ \Leftrightarrow \text{Wall distance }\
y^+=2^{-j^+} .~~~ %%
\EEA%%
This approach can be used to restore of spatial distribution of
various turbulent characteristics. In particular,  we can
understand $\ve^-_{j^+}$, presented in
Fig.~\ref{f:Vy-eps-y-profs}, as $\ve^-(y^+)$ with
$y^+=(2^{-j^+})$. As expected, the dissipation rate (normalized,
as in Fig.~\ref{f:Vy-eps-y-profs},   by \Re) is \Re-independent as
$\RE\to   \infty$. In this sense, the results in
Fig.~\ref{f:Vy-eps-y-profs} can be considered as universal. As
expected, at large $\RE$ the main energy dissipation occurs in a
narrow near-wall region, $y^+<40$. The dissipation rate at the
wall in our model is equal to%%
\bea\label{eq:ve}%%
 \ve|_\text{wall}=\lim_{j\to\infty}\hat\varepsilon^-_j/\RE \approx
0.21\,, \quad \RE> 10^4\,, %%
\EEA%%
 which is reasonably close to the result for smaller \Re,
 available in the DNS of Ref.~\cite{98MKM}: %%
\bea\label{eq:ve1}%%
\ve|_\text{wall}=0.166\,,\quad \RE\sim 200\ .%%
\EEA%%
Notice, the difference in \Re's and  that our  result~\Ref{ve} is
obtained in the simple model with just two adjustable parameters,
chosen to adjust very different characteristics of the flow, the
mean velocity profile. This allows one to consider the reasonable
correspondence of Eqs. \Ref{ve} and \Ref{ve1} as an argument in
favor of our simple MZS model.
%%
\subsubsection{\label{sss:glob}
$\RE$-dependence of the global flow characteristics }%%
Clearly, our approach  allows one to evaluate various global
characteristics of the turbulent wall bounded flows.

\begin{figure}%%
\epsfxsize=8.3cm \epsfbox{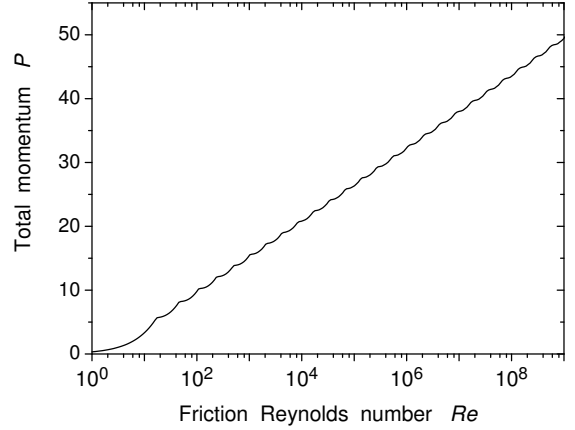}%%
\caption{\label{f:graph1}%%
Dependence of the total momentum of the flow $\mathcal P$ from the
Reynolds number $\RE$ in the channel flow.}
\end{figure}

The first example is the $\RE$-dependence of the  total momentum
(i.e. total flux) of the flow, shown  in Fig.~\ref{f:graph1}.
There is a laminar regime for $\RE<17 $, and a developed turbulent
regime for, say $\RE>100$. The ``period-2" oscillations are an
artifact of the model, caused by the  discretization with the
spacing parameter $\lambda=2$. These oscillations, however, are
small and should be ignored. In principle, they can be removed in
more ``advanced" versions of the MZS models with few variables in
each $j$-zone, $V_{j,\sigma}$, $u_{j,\sigma}$, responsible for
``$\sigma$-sub-zones of the $j$-zone.

%%
\begin{figure}%%
\epsfxsize=8.6cm \epsfbox{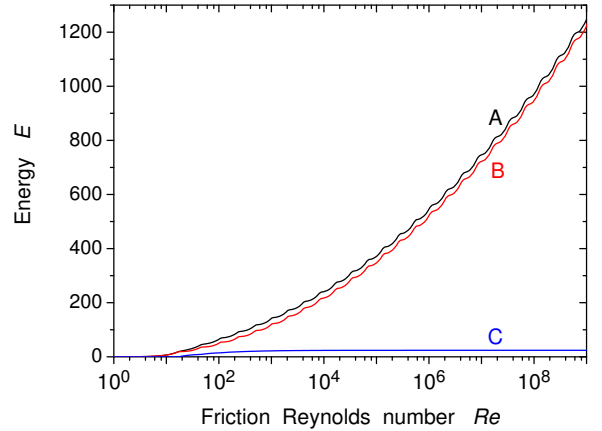}%%
\caption{\label{f:graph2}%%
Dependence of the total energy (A), the energy of the mean
$V$-subsystem (B), and the energy of the turbulent $u$-subsystem
(C) from the Reynolds number $\RE$. }
\end{figure}%%%%

The second example is the $\RE$-dependence of the total energy of
the system, as well as parts of the energy, containing in the mean
and turbulent subsystems, shown in Fig.~\ref{f:graph2}.

\begin{figure}%%
\epsfxsize=8.6cm \epsfbox{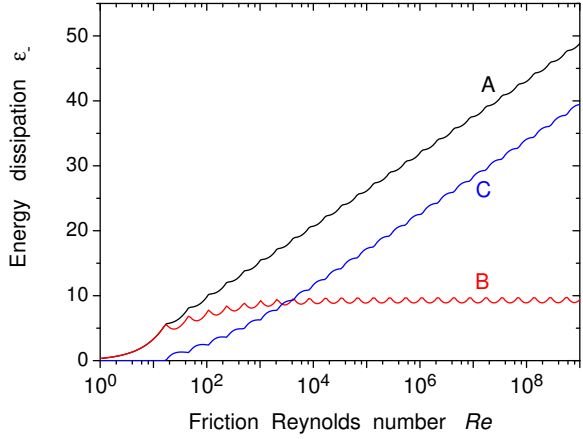}%%
\caption{\label{f:diss-Re}%%
Dependence of the total energy dissipation (A), energy dissipation
in  the mean $V$-subsystem (B), and energy dissipation in the
turbulent $u$-subsystem (C) from the Reynolds number $\RE$. }
\end{figure}%%
%%%%%%%%%%%%%%%%%%%%%%%%%%%%%%

Last, but not least, example is the $\RE$-dependence of the total
energy dissipation and that for the $V$- and $u$-subsystem, shown
in Fig.~\ref{f:diss-Re}. Total energy dissipation is equal to the
total energy influx, i.e. to the total momentum of the flow.
However, the distribution of energy dissipation between two
subsystems is very interesting.

One clearly sees  that some flow characteristics, like the energy
of turbulent subsystem and the energy dissipation in the mean flow
subsystem remain finite for vanishining  fluid viscosity, $\nu_0$.
In the same time, other characteristics, such as the total linear
momentum, the energy of the mean flow sub-system and the rate of
energy dissipation increase infinitely [like $\ln(1/\nu_0)$], i.e.
demonstrate a phenomenon of \emph{viscous anomaly}.  The MZS model
clearly demonstrate that the physical reason for that anomaly is
the separation in the physical space of the external forcing and
the friction: the external pressure gradient, that accelerate the
flow,  acts on the whole cross-section area of the flow, while the
friction force, that prevents the mean velocity from infinite
growth, acts only on the walls. To be able to maintain the
constant flux of the linear mechanical momentum toward the wall,
the amplitudes of the near-wall eddies of all scales, $u_j$, must
be $j$-independent. This immediately leads to a linear increase of
$V_j$ with $j$, decreasing from the viscous cutoff value, $j_0$,
toward the beginning of the cascade, $j=1$. Thus  the value of
$V_1$ is proportional to the total number of the cascade steps,
$j_0-1$. Decrease of $\nu_0$ to a half of its value adds one more
step in inertial interval of the momentum cascade. This leads to
an  increase of $V_1$  by $\sim \ln 2$. In the  dimensional units
this corresponds to the increase $V_1$ by $\sim V_\tau \ln 2\sim
\ln 2 \sqrt{L\nabla p}\ $.

\section{\label{s:sum}Summary}

We develop  a multi-zone shell (MZS) model for the wall-bounded
turbulent flows in the piecewise approximation, dividing
cross-section area into set of $N\sim\log_2 \RE$ $j$-zones. In
each zone the turbulence is assumed to be homogeneous and is
described in the framework of a shell model equation for the
``turbulent" shell velocities $u_{nj}(t)$. The mean velocity is
described by an additional set of $N$ zone variables $V_j(t)$,
that allow to reconstruct the mean velocity profile with the help
of a specially designed \PM-basis.

The MZS model

 \noindent\textbullet~conserves the actual integrals of motion of
the original NSE of the problem, \emph{energy, linear  and
angular momenta};

 \noindent\textbullet~respects  {Galilean } and  ``asymptotic"
{scale invariance}, NSE {type of nonlinearity}.

 \noindent\textbullet~in the relatively simple and analytically
transparent manner describes the basic physical phenomena in the
wall bounded flows for a huge interval of Reynolds numbers. They
include:
\begin{enumerate}
\item the  laminar velocity profile for {${\cal R}e<{\cal R}e_{\rm
cr}$}; %%
\item its instability at {${\cal R}e={\cal R}e_{\rm cr}$}, %%
\item intermediate, non-universal mean velocity profile at
moderate $\RE$;%%
\item universal profile for ${\cal R}e\gg {\cal R}e_{\rm cr}$ in
the viscous sublayer, buffer layer and log-law region;%%
\item spatial distribution of turbulent activity, of the rate of
energy dissipations, \emph{etc}.
\end{enumerate}
 \noindent\textbullet~allows additional adaptation  of the MZS
equations  for first few (energy containing) shells to the
particular flow geometries (like the channel, pipe, Couette flows,
\emph{etc.}) that may be based on the stability analysis of the
laminar regime or some other specific geometry determined
information.  This should improve description of the flow for
moderate $\RE < 1000$.

 \noindent\textbullet~may be generalized on the case of
 viscoelastic turbulent flows
(by adding additional shell variables for the polymeric
additives), on the case of particle laden suspensions, etc.%%
%%%%
%%
\begin{acknowledgements}
We thank Itamar Procaccia,  Nikolai   Nikitin  and Alex Yakhot for
useful discussions. The support of the Israel Science Foundation
governed by the Israeli Academy of Science is gratefully
acknowledged.
\end{acknowledgements}
\appendix

\section{\label{a:PM}%%
Equation for the laminar \PM-profile}%%
Here we present a proof that the laminar \PM-profile, $\B
V\SPM^0(\B \rho)$, given by \REF{Vy-lam}, satisfies \REF{NSE-2}.

To this end, we recall that the damping parameters $G_j\kappa_j^2$
are defined as matrix elements:
\begin{equation}
G_j\kappa_j^2 = -s_j^{-1}\left(\bm\Phi_j',\Delta\bm\Phi_j'\right)
\,.\end{equation} Substituting this definition into \Ref{Vj-lam},
\Ref{Vy-lam}, we obtain
\begin{eqnarray}
\bm V\SPM^0(\bm\rho) &=& -\RE\sum_j \frac{s_j\bm\Phi_j'(\bm\rho)}
{\left(\bm\Phi_j',\Delta\bm\Phi_j'\right)} \,.\end{eqnarray} Then,
the Laplacian of $\bm V\SPM^0$ is equal to
\begin{equation}\label{laplace}
\Delta\bm V\SPM^0(\bm\rho) = -\RE\sum_j
\frac{s_j\Delta\bm\Phi_j'(\bm\rho)}
{\left(\bm\Phi_j',\Delta\bm\Phi_j'\right)} \,.\end{equation}
Acting on this Laplacian with the \PM-projector \Ref{PM-proj}, we
obtain
\begin{eqnarray}\label{q-1}
  \hat{\bm P}\SPM\{\Delta\bm V\SPM^0\} =
  -\RE\sum_{jk} \frac{s_j}{s_k}\text{Re}\left\{ \bm\Phi_k
    \frac{\left(\bm\Phi_k\Delta\bm\Phi_j'\right)}
{\left(\bm\Phi_j',\Delta\bm\Phi_j'\right)}\right\}\,,%%
\end{eqnarray}
where one must distinguish the Reynolds number, $\RE$, from the
notation of the real part of something, Re$\{\dots\}$. Note that
by construction, different \PM-functions $\bm\Phi_j$ belong to the
different subspaces of the eigen-functions of the Laplace
operator. Also, the real $\bm\Phi_j'$ and imaginary $\bm\Phi_j''$
parts of these functions are orthogonal, and thus
\begin{equation}
\left(\bm\Phi_k,\Delta\bm\Phi_j'\right) =
\Delta_{kj}\left(\bm\Phi_j',\Delta\bm\Phi_j'\right)
\,.\end{equation} As a result, all non-diagonal terms in
(\ref{q-1}) vanish, and we obtain
\begin{eqnarray}\label{q-2}
  \hat{\bm P}\SPM\{\Delta\bm V\SPM^0\} &=&
  -\RE\sum_{j}\bm\Phi_j'
\,.\end{eqnarray}%%
Finally,  using property \Ref{unit-a}, we obtain
\begin{eqnarray}
  \hat{\bm P}\SPM\{\Delta\bm V\SPM^0\}=
   -\RE\,{\bf\hat x}
\,,\end{eqnarray} %%
which actually is  \REF{NSE-2}.%%

\section{\label{a:iter}%%
Iteration procedure for solving MZS equations in the effective
turbulent damping approximation} %%
Consider  the MZS-Eqs.\Ref{turb-damp} for some large $\RE$.  As we
discussed, there are two regions of $j$, namely  $j<j\sb{max}$ and
$j\ge j\sb{max}$. In the ``turbulent region", $j< j\sb{max}$, the
laminar solution, $u_j=0$,  is unstable with respect to excitation
of turbulent amplitudes $u_j$ and thus, in the stationary regime,
$u_j\ne 0$. For $d>0$,  without loss of generality all $u_j$ can
be taken real and positive definite. In the turbulent region the
stationary velocities $V_j$ and $u_j$ satisfy the equation:%%
\begin{subequations}\label{eq:stT}
\begin{eqnarray}\label{eq:stT-a}
\Gamma_jV_j &=& 1 + d\,\kappa_j(u_{j-1}^2-u_j^2) \,,\\
\label{eq:stT-b} %%
\gamma_j +\alpha\kappa_j u_j&=&\frac{d}{2
\sigma_j}(V_j-V_{j+1})\,,\br %%
u_j&\ge 0& \,,\quad j < j\sb{max}  \ .%%
\end{eqnarray} \end{subequations}%%
In the ``laminar region", $j \ge j\sb{max}$:%%
\bea\label{eq:stL}%%
V_j=1/\Gamma_j\,,\quad u_j=0\,, \quad j\ge j\sb{max}\ .%%
\EEA %%
One can substitute $V_j$ from \REF{stT-a} to \REF{stT-b} and get
an equation, connecting the triad $u_{j-1}$, $u_j$, and $u_{j+1}$.
Unfortunately, a direct iteration procedure in this
``straightforward" equation is unstable and do not converge to the
stationary solution.

To find  the stable, stationary solution of Eqs. (\ref{eq:stT})
numericcaly, we develop a stable iteration procedure, which is
based on the physical understanding of these equations as
describing the momentum flux from $j=1$ toward large $j$ in the
``interaction triads"  $ V_j$, $V_{j+1}$ and $u_j$. Denote as
$V_j^{(p)}$, $\tilde{V}_{j+1} ^{(p)}$ and $u_j^{(p)}$ the solution
of following three algebraic equations on the $j$-step in the
$p$-th iteration run:%%
\begin{eqnarray}\nn
\Gamma_jV_j^{(p)} &=& 1 + d\,\kappa_j\{ [u_{j-1}^{(p)}]^2
-[u_j^{(p)}]^2 \} \,,\br %%
\Gamma_{j+1}\tilde{V}_{j+1}^{(p)} &=& 1 +
d\,\kappa_{j+1}\{[u_{j}^{(p)}]^2 -[u_{j+1}^{(p-1)}]^2\} \,,\\
\label{eq:iter1}%%
\gamma_j +\alpha\kappa_j u_j^{(p)}&=&\frac{d}{2
\sigma_j}[V_j^{(p)}-\tilde{V}_{j+1}^{(p)}]\,,\br u_j^{(p)}&\ge&
0\,,\quad\text{for}\  j\sb{max}>j\ge 1\,, \quad  u_0^{(p)} = 0%%
\ .%%%%
\end{eqnarray}
As the ``initial condition" at $p=0$ we take  the (unstable)
laminar solution:
\[
V_j^{(0)} = \Gamma_j^{-1} \,,\quad u_j^{(0)}=0\ .%%
\]%%

In the first step, $j=1$, of each iteration run  one takes
$u_0^{(p)}=0$. Finding $V_1^{(p)}$,  $\tilde{V}_2^{(p)}$, and
$u_1^{(p)}$ one takes in \REF{iter1} $j=2$, finds $V_2^{(p)}$,
$\tilde{V}_3^{(p)}$ and $u_2^{(p)}$ and so on until on some $j_0$
step one gets negative (or complex) solution for $u_{j_0}^{(p)}$.
It means that this amplitude is stable and has to be taken zero,
$u_{j_0}^{(p)}=0$. Accordingly, $j_0=j\sb{max}$. For all
$j>j\sb{max}$ one takes the laminar solution:
$V_j^{(p)}=1/\Gamma_j$, $u_j^{(p)}=0$.

After that  one begin the next, $p+1$, iteration run, starting
again from its first step, $j=1$. It can be shown, that the
velocities $V_1^{(p)}$ form monotonically decreasing sequence with
increasing $p$ and are always positive. Since a limited from
below, monotonically decreasing sequence always have some finite
limit, this proves the convergence and stability of our iteration
scheme. The calculations show, that for Reynolds numbers
$\RE\le10^9$ velocities $V_j$ and $u_j$ converge (with accuracy
about $10^{-6}$) after $100-200$ iterations runs.

%%%%%%%%%%%%%%%%%

\end{document}